\documentclass[
reprint,
 showkeys,
 amsmath,amsmath,
aps,
pra,
floatfix,
]{revtex4-1}
\usepackage{graphicx}
\usepackage{bm}
\usepackage[breaklinks=true,colorlinks=true,linkcolor=blue,urlcolor=blue,citecolor=red]{hyperref} 
\usepackage{cleveref}
\crefname{equation}{Eq.}{Eqs.}
\crefname{section}{Sec.}{ Secs.}
\crefname{figure}{Fig.}{Figs.}
\crefname{tabular}{Tab.}{Tabs.}
\creflabelformat{equation}{#2\textup{(#1)}#3}
\begin{document}
\title{Steady-state mechanical squeezing and ground-state cooling of a Duffing anharmonic oscillator
in an optomechanical cavity assisted by a nonlinear medium }

\author{F. Momeni$^1$}
\email{farmomeni.1392@gmail.com}
\author{M. H. Naderi$^{1,2}$}\email{mhnaderi@sci.ui.ac.ir}%
 \affiliation{$^1$Department of Physics, Faculty of Science, University of Isfahan, Hezar Jerib, 81746-73441, Isfahan, Iran\\
$^2$ Quantum Optics Group, Department of Physics, Faculty of Science, University of Isfahan, Hezar Jerib, 81746-73441, Isfahan, Iran
}%

\date{\today}

\begin{abstract}
In this paper, we study theoretically a hybrid optomechanical system consisting of a degenerate optical parametric amplifier inside a driven optical cavity with a moving end mirror which is modeled as a stiffening  Duffing-like anharmonic quantum mechanical oscillator. By providing analytical expressions for the critical values of the system parameters corresponding to the emergence of the multistability behavior in the steady-state response  of the system, we show that the stiffening mechanical Duffing anharmonicity reduces the width of the multistability region while the optical parametric nonlinearity can be exploited to drive the system toward the multistability region. We also show that for appropriate values of the mechanical anharmonicity strength the steady-state mechanical squeezing and the ground-state cooling of the mechanical resonator can be achieved. Moreover, we find that the presence of the  nonlinear gain medium can lead to the improvement of the mechanical anharmonicity-induced cooling of the mechanical motion, as well as to the mechanical squeezing beyond the standard quantum limit of 3 dB.
\end{abstract}
\keywords{nonlinear cavity optomechanics, Duffing anharmonicity, optical parametric amplifier, multistability, mechanical motion cooling, mechanical squeezing}
\maketitle


\section{\label{sec:int}Introduction}
In recent years, cavity optomechanical systems in which the radiation pressure force induces a coupling between the radiation field of a high-finesse cavity and the mechanical motion of a movable mirror have attracted much research attention~\cite{o1,o2,o3,o4,o5,o6,o7}. The growing interest in such systems is associated with their applications in a wide range of research topics including the generation of optomechanical entanglement~\cite{o8,o9,o10}, the ground-state cooling of the mechanical mode~\cite{o11,o12,o13,o14}, detection and interferometry of gravitational waves~\cite{o15}, position or force sensing~\cite{o16,o17,o18,o19}, optomechanically induced transparency realization~\cite{o20}, coherent state transfer between cavity and mechanical modes~\cite{o21}, and generation of nonclassical states of the mechanical and optical modes~\cite{o22,o23,o24}. Most of these applications are based on the intrinsic  nonlinear nature of the radiation pressure interaction. This nonlinearity is due to the fact that in a typical cavity optomechanical system the position of the mechanical oscillator modulates the resonance frequency of the cavity mode. In other words, the optical length of the cavity depends on the intensity of the intracavity field, and consequently the optomechanical cavity behaves effectively as a rigid cavity filled with a nonlinear Kerr medium~\cite{o25,o26}. The intrinsic optomechanical nonlinearity can be identified by the optomechanically induced transparency~\cite{o27}.

In addition to the inherent nonlinearity, the  cavity optomechanical systems can contain two types of nonlinearity, one of which is associated to the cavity field and the other one to the mechanical oscillator. In recent years, there has been a growing interest in nonlinear optomechanical cavity systems in which the nonlinearity is mainly contributed by  nonlinear media such as the optical Kerr medium~\cite{o28}, the optical parametric amplifier (OPA)~\cite{o29,o30}, or a combination of both of these nonlinear media (Kerr-down conversion nonlinearity)~\cite{o31,o32,o33}. In fact, the idea of combining nonlinear optics and optomechanics, with the aim of enhancement of quantum effects, has resulted in some interesting physical phenomena. It has been shown~\cite{o28} that a Kerr nonlinear medium inside an optomechanical cavity inhibits the normal-mode splitting due to the photon blockade mechanism, reduces the photon number fluctuation, and provides a coherently controlled dynamics for the moving mirror, which further could be useful in the realization of tuneable quantum-mechanical devices in the future. On the other hand, when an optomechanical cavity contains an OPA the cooling of the mechanical motion and the normal-mode splitting can be considerably improved due to the significant enhancement of the optomechanical coupling strength~\cite{o29,o30}.  It has also been demonstrated ~\cite{o34} that the squeezing of the cavity field generated by an OPA placed inside an optomechanical cavity can be transferred to the movable mirror with high efficiency in the resolved sideband limit. Moreover, it has been shown that the OPA can give rise to the improvement of the entanglement between one cavity mode and one mechanical mode~\cite{o35}, between two cavity modes which jointly interact with a mechanical resonator~\cite{o36}, between two mechanical modes of two coupled optomechanical cavities~\cite{o37}, and between multi-cavity and multi-mechanical modes~\cite{o38}. In addition, the manipulation of the optomechanically induced transparency behavior and coherent control of the entanglement between the vibrational modes of the micromirros in an optomechanical cavity made by two movable mirrors which contain a Kerr-down-conversion nonlinear crystal have been studied in Ref.~\cite{o31} and Ref.~\cite{o33}, respectively. 

In all of the above-mentioned investigations the mechanical resonator has been treated as a pure quantum harmonic oscillator.  It is important to mention that the dynamics of a purely harmonic mechanical mode is analogous to its classical counterpart, in the sense that the  expectation values of the canonical observables obey the classical equations of motion~\cite{o39}. Therefore, in order to detect the quantum behavior of the mechanical mode, introducing an additional nonlinearity (or \textit{anharmonicity}) may be useful. For micro- and nano-mechanical resonators in the sub-gigahertz range, the intrinsic (geometrical) nonlinearity is usually very weak with nonlinear amplitude smaller than $10^{-15}\omega_m$ ($\omega_m$ being the mechanical frequency)~\cite{o40}, and thus its contribution is relevant only in the regime of large oscillation amplitudes. Several schemes have been proposed to generate strong mechanical anharmonicity in the quantum regime. For example, in Ref.~\cite{o41} the authors have proposed a scheme based on subjecting a namomechanical resonator to inhomogeneous external electrostatic fields. This procedure can effectively reduce the resonator's stiffness, and consequently its resonance frequencies, which in turn enhances the amplitude of the oscillator's zero-point motion, leading to an amplification of its nonlinearity per phonon. The geometrical nonlinearities can be either stiffening (with positive parameter of nonlinearity) or softening (with negative parameter of nonlinearity). For some nanodevices, such as cantilever nanobeam, both softening and stiffening nonlinearities have been experimentally observed~\cite{o42,o43}. Stiffening geometrical nonlinearity has been used for quantum control and quantum information processing \cite{o41,o44} as well as for generating steady-state mechanical squeezing in optomechanical systems~\cite{o45}, whereas it has been shown~\cite{o46,o47} that softening nonlinearity can be a limiting factor for  mechanical cooling and squeezing. It has been found~\cite{o48} that geometrical nonlinearity can be exploited to generate robust steady-state optomechanical entanglement.
In Ref.~\cite{o49} the authors have proposed a method to engineer giant nonlinearities in a mesoscopic quantum resonator by using a simple auxiliary system perturbatively coupled to the resonator. Inspired by this method, a theoretical scheme~\cite{o45} has been presented to generate strong steady-state mechanical squeezing in an optomechanical system via cavity cooling and a quartic nonlinearity in the displacement of the mechanical oscillator (known as the Duffing nonlinearity~\cite{o50}) which is achieved by coupling the mechanical oscillator to an auxiliary highly nonlinear system, such as an external electrode or a qubit. The generation of steady-state mechanical squeezing via engineering the Duffing nonlinerity in a double-cavity optomechanical system \cite{o51} or by engineering a cubic mechanical nonlinearity in a hybrid atom-optomechanical system~\cite{o52} has also been theoretically studied. 

Motivated by the above-mentioned interesting features of nonlinear optomechanical cavity systems, in this paper we aim to study an optomechanical cavity which contains both mechanical and optical nonlinearities, i.e., a driven hybrid optomechanical cavity with a Duffing-like movable mirror that contains a degenerate OPA. By investigating the roles of the Duffing anharmonicity and the gain nonlinearity in the emergence of multistability  behavior of the system, we show that the stiffening Duffing anharmonicity greatly affects the width of the multistability region, though the OPA manifests its role in driving the system toward multistability region. In addition, it is shown that whereas the mechanical anharmonicity suppresses the amplitude of mechanical oscillations, it has no significant effect on the intracavity intensity. We also explore the effect of the Duffing anharmonicity on the ground-state cooling of the mechanical motion as well as on the mechanical quadrature squeezing. We find that in the absence of the   nonlinear gain medium the mechanical mode can be cooled down if the Duffing anharmonicity is not so strong, while strong mechanical squeezing can be achieved as the mechanical nonlinearity becomes stronger. On the other hand, the results reveal that in the  presence of the   nonlinear gain medium the mechanical anharmonicity-induced cooling of the mechanical motion as well as the mechanical squeezing can be enhanced.

The paper is structured as follows. In \cref{sec:2nd}, we describe the physical model of the system under consideration, give the quantum Langevin equations, and analyze the stability of the mean-field solutions as well as the dynamics of quantum fluctuations. In \cref{sec:3th}, we investigate the effects of both the mechanical anharmonicity and the gain nonlinearity on the ground- state cooling and quadrature squeezing of the mechanical oscillator. We summarize our conclusions in \cref{sec:4th}. In addition, we derive the critical values of the system parameters corresponding to the emergence of the multistable behavior in the  Appendix.

\section{\label{sec:2nd}Theoretical description of the system}
\begin{figure}[ht]
\centering
\includegraphics[scale=0.52]{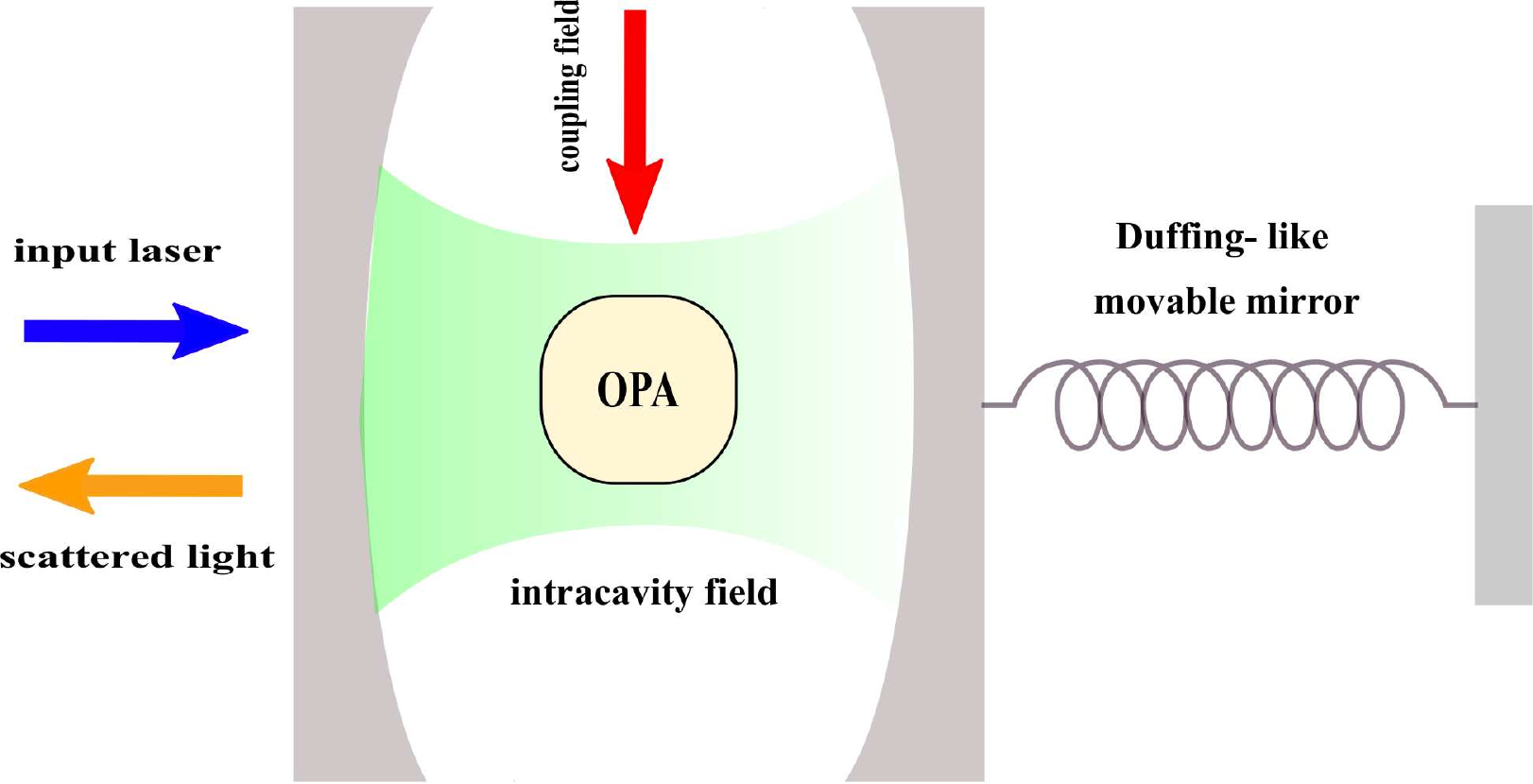}
\caption{\label{fig1}Schematic illustration of a hybrid optomechanical system consisting of a Fabry-Perot cavity with one fixed mirror and one movable mirror modeled as a nonlinear Duffing-like oscillator. The cavity mode is coherently driven by an input laser through the fixed mirror. A degenerate OPA is placed inside the cavity that is pumped by a coupling field to produce
parametric amplification.}
\end{figure}

As depicted in \cref{fig1}, we consider a nonlinear optomechanical cavity composed of a degenerate OPA placed inside a single-mode Fabry-Perot cavity formed by a fixed partially transmitting mirror and one movable perfectly reflecting mirror in equilibrium with its environement at a low temperature. The movable mirror is free to move along the cavity axis and is treated as a Duffing-like quantum mechanical oscillator with effective mass $m$, frequency $\omega_m$, energy decay rate $\gamma_m=\frac{\omega_m}{Q_m}$ ($Q_m$ being the mechanical quality factor), and Duffing nonlinearity parameter $\lambda$. The cavity field is coherently driven by an input monochromatic laser field with frequency $\omega_L$ and amplitude $|\varepsilon|=\sqrt{\frac{2\kappa_c  P_{in}}{\hbar \omega_L}}$ through the fixed mirror ($P_{in}$  is the input  laser power and $\kappa_c $ is the cavity decay rate). In addition, the system is pumped by a coupling field to produce parametric oscillation in the cavity. We restrict the model under consideration to the case of single-cavity and mechanical modes. The single mode description for both cavity field and mirror motion is valid whenever scattering of photons from the driven mode into other cavity modes is negligible~\cite{o53}, and if the detection bandwidth is chosen such that it includes only a single, isolated, mechanical resonance and mode-mode coupling is negligible~\cite{o54}.

The total Hamiltonian of the system in a frame rotating at the laser frequency $\omega_L$ is given by
\begin{align}
&\hat{H}=\hbar \Delta \;\hat{a}^\dagger \hat{a}+\hbar\omega_m \hat{b}^\dagger\hat{b}+\hbar\frac{\lambda}{2}(\hat{b}+\hat{b}^\dagger)^4 - \hbar g \hat{a}^\dagger \hat{a}(\hat{b}+\hat{b}^\dagger) \nonumber \\ & \hspace*{1cm}+i \hbar G_0\,(\hat{a}^{\dagger^2 }\,e^{i \theta} - \hat{a}^2\,e^{-i \theta}) + i \hbar (\varepsilon \hat{a}^\dagger - \varepsilon^*\hat{a})\label{Ham},
\end{align}
where $\Delta=\omega_c-\omega_L$ is the cavity detuning from the frequency of the driving laser. In the above Hamiltonain, the first term denotes the cavity mode energy (described by the photon creation and annihilation operators $\hat{a}^\dagger$ and $\hat{a}$), and the second and third terms account for the Hamiltonian of the nonlinear mechanical mode (described by the phonon creation and annihilation operators  $\hat{b}^\dagger$ and $\hat{b}$ ) with Duffing nonlinearity (anharmonicity) strength $\lambda>0$. As mentioned in the Introduction, the intrinsic anharmonicity of the sub-gigahertz micro- and nano-mechanical resonators is usually very weak in the regime of very small oscillation amplitudes. However, one can obtain a strong nonlinearity  through coupling the mechanical mode to an ancilla system, e.g., an external qubit~\cite{o49}. The fourth term in the Hamiltonian of \cref{Ham} describes the optomechanical interaction between the cavity field and the mechanical oscillator via the radiation pressure force with single-photon coupling strength $g=\frac{\omega_c}{L}\sqrt{\frac{\hbar}{2 m \omega_m}}$ ($L$ being the cavity length in mechanical equilibrium). The fifth term corresponds to the coupling of the intracavity field with the OPA; $G_0$ is the nonlinear gain which is proportional to the pump power driving the OPA, and $\theta$ is the phase of the field which drives the OPA. Finally, the last term describes the coupling between the cavity mode and the input  laser. 

The dynamics of the optomechanical system is fully characterized by the quantum Langevin equations obtained by adding the corresponding damping and input noise terms to the Heisenberg equations associated with  the Hamiltonian of \cref{Ham}, 
\begin{subequations}\label{qle}
\begin{align}
&\dot{ \hat{a}}=-i\Delta \hat{a} + i g \hat{a}(\hat{b}+\hat{b}^\dagger)+2G_0 e^{i\theta}\hat{a}^\dagger \nonumber\\  &\hspace*{4cm}- \kappa_c\, \hat{a}+\varepsilon +\sqrt{2\kappa_c}\,\hat{a}_{in},\\& \dot{ \hat{b}}=-i\omega_m \hat{b} -2i \lambda (\hat{b}+\hat{b}^\dagger)^3 + i g \hat{a}^\dagger \hat{a}\nonumber\\  &\hspace*{4cm} - \gamma_m \,\hat{b}+\sqrt{2\gamma_m} \,\hat{b}_{in},\label{beq}
\end{align}
\end{subequations}
where the cavity input noise  $\hat{a}_{in}$ and the input thermal noise of the mechanical oscillator $\hat{b}_{in}$, with zero mean values, satisfy the commutation relation $[\hat{a}_{in}(t), \hat{a}_{in}^\dagger(t')]=[\hat{b}_{in}(t), \hat{b}_{in}^\dagger(t')]=\delta(t-t')$. The input noise operator $\hat{a}_{in}$ satisfies the Markovian correlation functions, i.e., $\langle \hat{a}_{in}(t)\hat{a}_{in}^\dagger(t')\rangle=(1+\bar{n}_{ph})\delta(t-t')$, $\langle \hat{a}_{in}^\dagger(t)\hat{a}_{in}(t')\rangle=\bar{n}_{ph}\delta(t-t')$, $\langle \hat{a}_{in}^\dagger(t)\hat{a}_{in}^\dagger(t')\rangle=\langle \hat{a}_{in}(t)\hat{a}_{in}(t')\rangle=0$ with the average thermal photon number $\bar{n}_{ph}$ which is nearly zero at optical frequencies \cite{o55}. In addition, in the limit of high mechanical quality factor ($Q_m\gg 1$) the mirror Brownian thermal noise $\hat{b}_{in}$ can be faithfully considered as a Markovian noise \cite{o56} whose nonvanishing correlation functions are given by 
\begin{align}
&\langle \hat{b}_{in}(t)\,\hat{b}_{in}^\dagger(t')\rangle =(1+\bar{n}_m)\delta(t-t'),\nonumber\\& \langle\hat{b}_{in}^\dagger(t)\hat{b}_{in}(t'
)\rangle=\bar{n}_m \,\,\delta(t-t'),
\end{align}
where $\bar{n}_m=[\text{exp}(\hbar \omega_m/k_B T)-1]^{-1}$ is the mean number of thermal phonons in the absence of the optomechanical coupling with $k_B$ and $T$ being the Boltzmann constant and the temperature of the mechanical bath, respectively.   

    Analyzing the quantum dynamics of the full nonlinear system described by the coupled nonlinear operator equations of motion (\ref{qle}) is a hard task. In order to solve analytically these equations, one can adopt the standard linearization procedure \cite{o6} in which both the cavity and mechanical modes are split into a steady-state mean value and a zero-mean quantum fluctuation, i.e., $\hat{O}=\langle \hat{O}\rangle_s\,+\,\delta \hat{O}$ with  $\langle \delta \hat{O}^\dagger\delta \hat{O}\rangle_s/ \langle \hat{O}^\dagger \hat{O}\rangle_s \ll 1\,\,(\hat{O}=\hat{a}, \hat{b})$. In this way, a set of nonlinear algebraic equations for the mean-field values and another set of linear ordinary differential equations for the quantum fluctuations will be obtained. However, a remark on the validity of the linearization approximation adopted for the system under consideration is in order. In the standard optomechanical systems where the nonlinearity is only due to the radiation pressure coupling, the linearization approximation is reliable if the cavity is intensely driven so that the intracavity field is strong \cite{o56}. Nevertheless, the validity of this approximation in the present optomechanical system in which there exist two additional types of nonlinearity (parametric amplification and Duffing nonlinearities) should be reexamined. This will be detailed in the following.
\subsection{\label{subsec:A}The mean-field solutions and their stability }
Under the mean-field approximation~\cite{o57}, i.e., $\langle\hat{a}^\dagger \hat{a}\rangle \approx \langle\hat{a}^\dagger\rangle\langle \hat{a}\rangle $ and $\langle\hat{a}\hat{b}\rangle \approx \langle\hat{a}\rangle\langle \hat{b}\rangle $, which is applicable when the coupling between the cavity and mechanical modes is weak, the mean-value equations read as
\begin{subequations}\label{mqle}
\begin{align}
&\hspace*{-0.2cm} \langle\dot{\hat{a}}\rangle =-i\Delta \langle\hat{a}\rangle + i g\langle\hat{a}\rangle\langle\hat{b}+\hat{b}^\dagger\rangle +2G_0 e^{i\theta}\langle\hat{a}^\dagger\rangle +\varepsilon - \kappa_c\, \langle\hat{a}\rangle, \label{mqlea}\\&\hspace*{-0.2cm}  \langle\dot{\hat{b}}\rangle =-i\omega_m \langle\hat{b}\rangle -2i \lambda \langle(\hat{b}+\hat{b}^\dagger)^3\rangle + i g \langle\hat{a}^\dagger\rangle\langle\hat{a}\rangle - \gamma_m \,\langle\hat{b}\rangle. \label{mqleb}
\end{align}
\end{subequations}
To evaluae $\langle(\hat{b}+\hat{b}^\dagger)^3\rangle$ in \cref{mqleb}, we write the cubic term $(\hat{b}+\hat{b}^\dagger)^3$ as~\cite{o58}
\begin{equation}
(\hat{b}+\hat{b}^\dagger)^3=:(\hat{b}+\hat{b}^\dagger)^3: +\, 3 \, (\hat{b}+\hat{b}^\dagger),
\end{equation}
in which the symbol $:\,:$ refers to normal ordering. Furthermore, in semiclassical approximation  we can write $\langle:(\hat{b}+\hat{b}^\dagger)^3:\rangle=\langle(\hat{b}+\hat{b}^\dagger)\rangle^3$. Assuming $\gamma_m\ll \omega_m,\kappa_c,\lambda$, the steady- state solutions of \cref{mqlea,mqleb} read as follows\begin{subequations}\label{ss}
\begin{align}
&\langle\hat{a}\rangle_{s} =\frac{(\kappa_c-i\Delta')\varepsilon +2G_0 e^{i\theta}\varepsilon^*}{\Delta{'}^{2}+\kappa_c^2-4 G_0^2},\label{as}\\&16\lambda \langle\hat{b}\rangle_{s}  ^3+(\omega_m\,+\,12\lambda )\langle\hat{b}\rangle_{s} - g \langle\hat{a}^\dagger\rangle_{s} \langle\hat{a}\rangle_{s} =0,\label{bs}
\end{align}
\end{subequations}
where $\Delta'=\Delta-2g \langle\hat{b}\rangle_{s}$ is the effective detuning of the cavity which includes the effects of the nonlinearities of the system. Without loss of generality, $\langle\hat{a}\rangle_{s}$ can be taken real and positive by an appropriate choice of the phase of $\varepsilon$ so that,
\begin{equation}\label{aas}
\langle\hat{a}\rangle_{s} =\frac{|\varepsilon|}{\sqrt{(\Delta' - 2 G_0 \sin \theta )^2+(\kappa_c - 2G_0 \cos \theta )^2}}.
\end{equation}
In the absence of both the Duffing anharmonicity and gain nonlinearity ($\lambda,\,G_0=0$), the steady-state mean number of intracavity photons determined by $I_a=|\langle\hat{a}\rangle_{s}|^2=\frac{|\varepsilon|^2}{\Delta{'}^{2}+\kappa_c^2}$, satisfies a third-order equation with three real solutions at most two of which are dynamically stable~\cite{o32}. However, when $G_0$ is comparable to $\kappa_c$, $I_a$ obeys a fifth-order equation that can have at most five real roots, three of which are dynamically stable~\cite{o59}. The existence of multistability in the behavior of the system depends on the input laser power $P_{in}$, bare detuning $\Delta$, steady-state mechanical oscillation amplitude $\beta_{s}=\langle\hat{b}\rangle_s$ (and therefore the intensity of the intracavity field), and their corresponding critical values. For the system under consideration, with the assumption $0\leq\lambda/\omega_m\ll 1$, the critical values are given by (see the Appendix for details)
\begin{subequations}
\begin{align}
\label{criteqs1}&\beta_{s}^{crit}=\bar{k}\Big(1+64\bar{k}^2\frac{\lambda }{\omega_m}+256\bar{k}^2\frac{\lambda^2}{\omega_m^2}(16\bar{k}^2-1)\Big)^{1/2},\\\label{criteqs2}& \Delta^{crit}\,=2G_0\,\sin \theta \,+\,4g\bar{k}\Big(1+16\bar{k}^2\frac{\lambda }{\omega_m}\nonumber\\&\hspace*{4cm}+192\bar{k}^2\frac{\lambda^2 }{\omega_m^2} (4 \bar{k}^2-1)\Big),\\&\label{criteqs3} P_{in}^{crit}=\frac{4 \hbar g\,\bar{k}^3\,  \omega_L \omega_m}{\kappa_c}\Big(1+12\frac{\lambda }{\omega_m}(1+4\bar{k}^2)\nonumber\\&\hspace*{5cm}+3072\,\bar{k}^4\frac{\lambda^2 }{\omega_m^2}\Big),
\end{align}\end{subequations}
where $\bar{k}=|(\kappa_c\,-\,2G_0\,\cos \theta)/2g|$ is  a measure of the strength of the single-photon optomechanical coupling and can be controlled via the properties of  the  nolinear gain medium. The system enters the multistability region whenever the conditions $\beta_s>\beta_s^{crit}$, $\Delta>\Delta^{crit}$, and $P_{in}>P_{in}^{crit}$ are fulfilled simultaneously. In the absence of the mechanical anharmonicity ($\lambda=0$), Eq. (\ref{aas}) together with critical values given by Eqs.~(\ref{criteqs1}-\ref{criteqs3}) clearly show that the OPA plays an essential role in driving the system into or removing it from the multistable region by changing the critical quantities of the system. For $\cos\theta>0$ and $2G_0<\kappa_c/\cos\theta$, the OPA simultaneously reduces the critical power and increases the intracavity intensity, causing the system to approach the multistable region. On the other hand, for $\cos\theta<0$ and an arbitrary value of the nonlinear gain $G_0$, due to increasing the critical power and reducing the intracavity field, the OPA prevents the appearance of the multistability. 

\begin{figure}[ht!]
  \centering
   \includegraphics[scale=0.39]{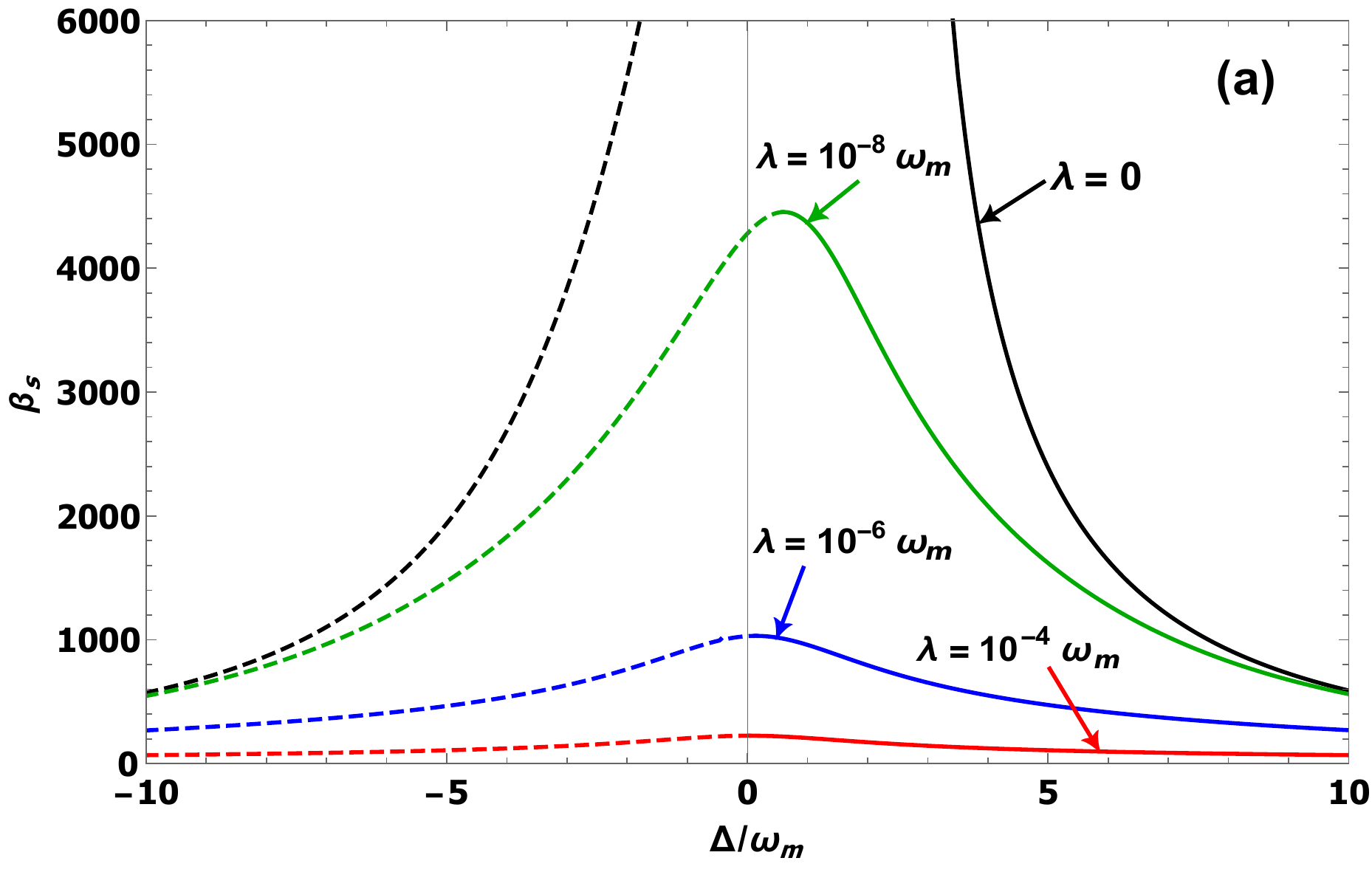}  \par
   \includegraphics[scale=0.37]{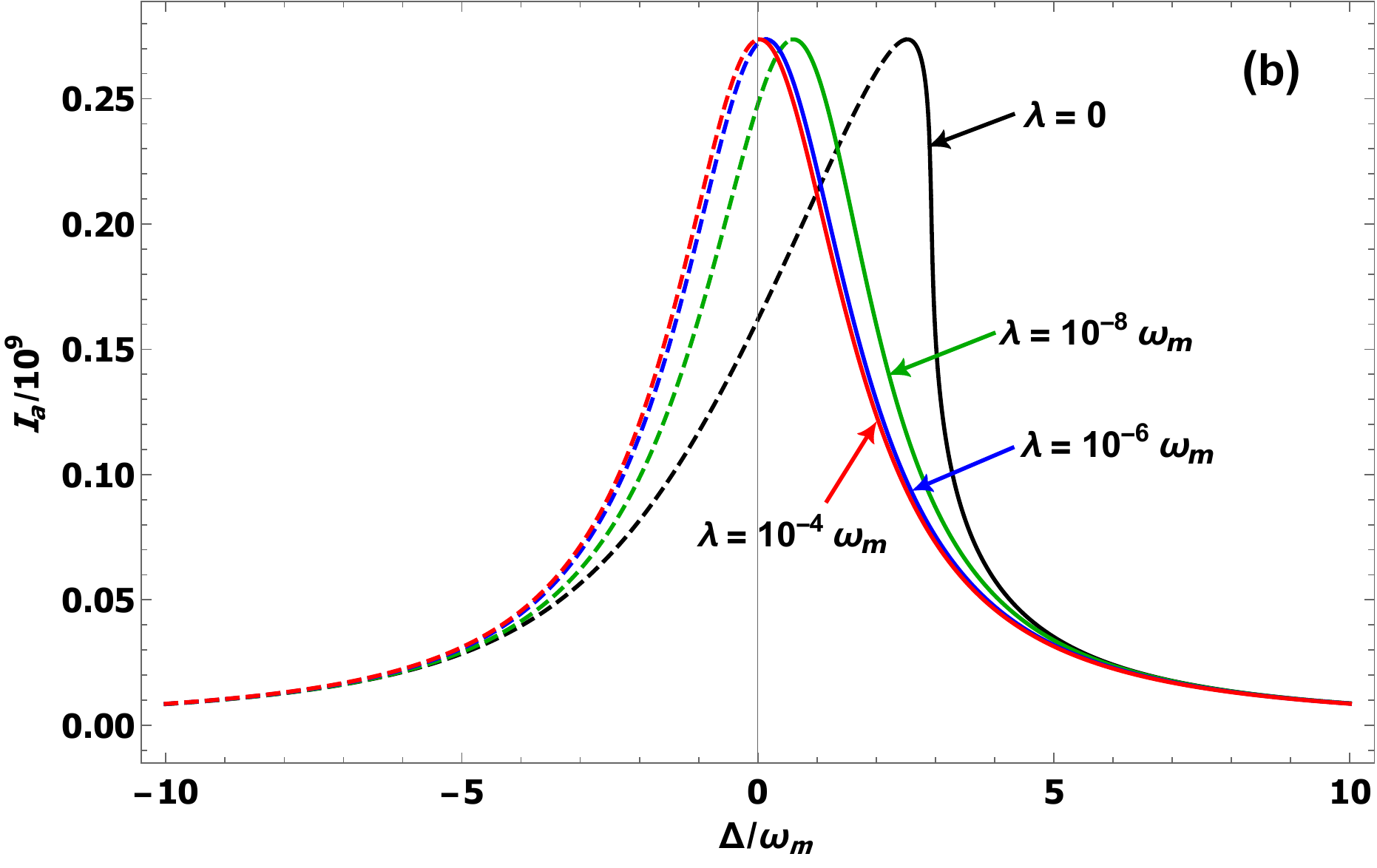}
    \caption{\label{fig2}(a) Steady-state mechanical oscillation amplitude and (b) steady-state intracavity intensity versus the normalized bare detuning $\Delta/\omega_m$ for a bare cavity ($G_0=0$) and for different values of the Duffing nonlinearity strength. Here, we have used the following set of experimentally realizable parameters \cite{o60,o61,o62,o63}: Length of cavity $L=1\,\text{mm}$, driving laser wavelength $\lambda_L=512\,\text{nm}$, input laser power $P_{in}=3\, \text{mW}$, cavity finesse $\mathcal{F}=1.67*10^4$ (corresponding to $\kappa_c \approx 0.9 \omega_m$), effective mass $m=5\, \text{ng}$, mechanical resonance frequency $\frac{\omega_m}{2\pi}=5\, \text{MHz}$, and mechanical quality factor $Q_m=10^5$. In this figure and the subsequent figures in this section the unstable steady-state solutions are represented by dashed lines. The stability conditions can be derived by applying the Routh-Hurwitz criterion (see Eq.~(\ref{RH})). }
\end{figure}

\begin{figure}[h!]
 \centering
   \includegraphics[scale=0.44]{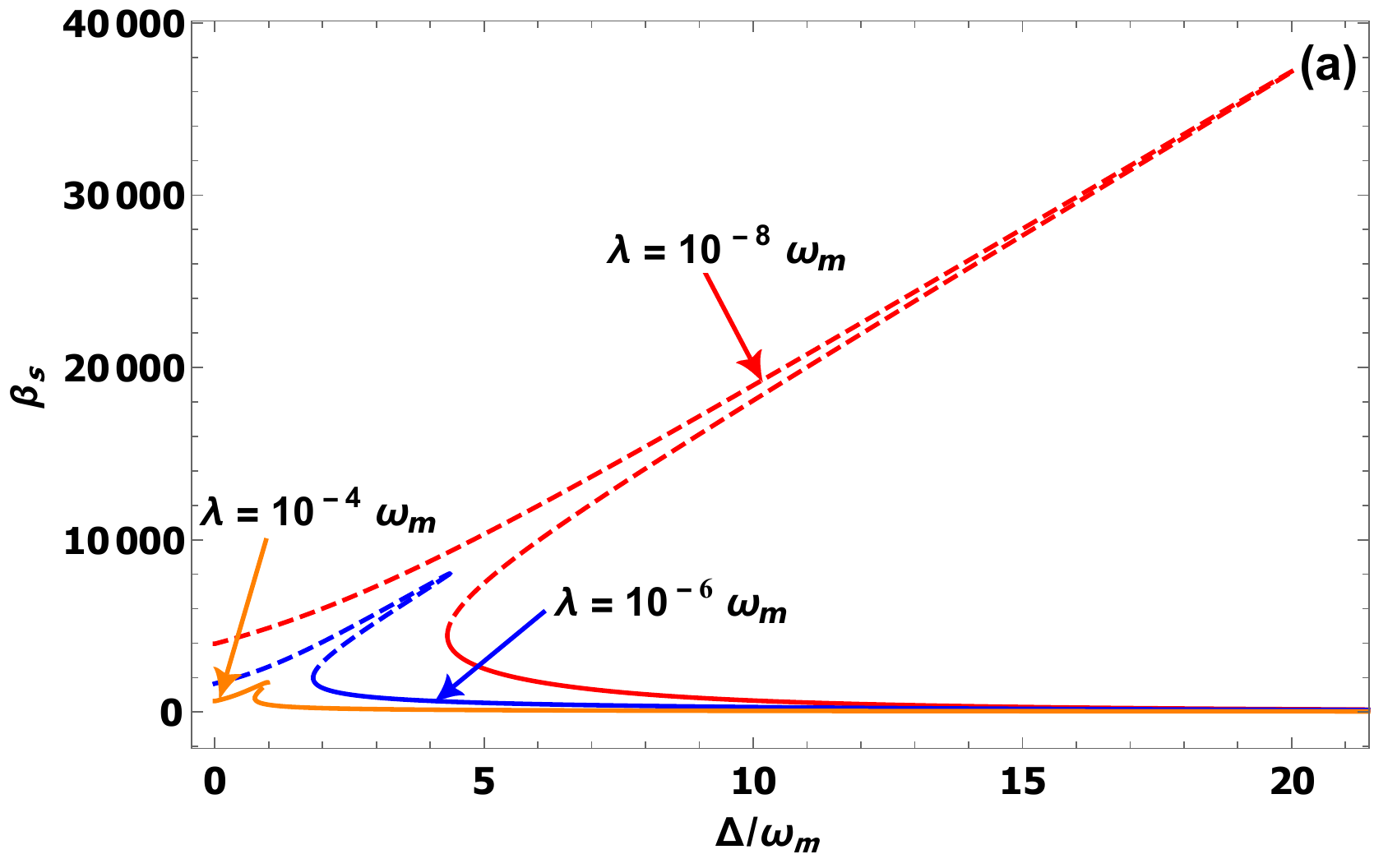} \par
  \includegraphics[scale=0.42]{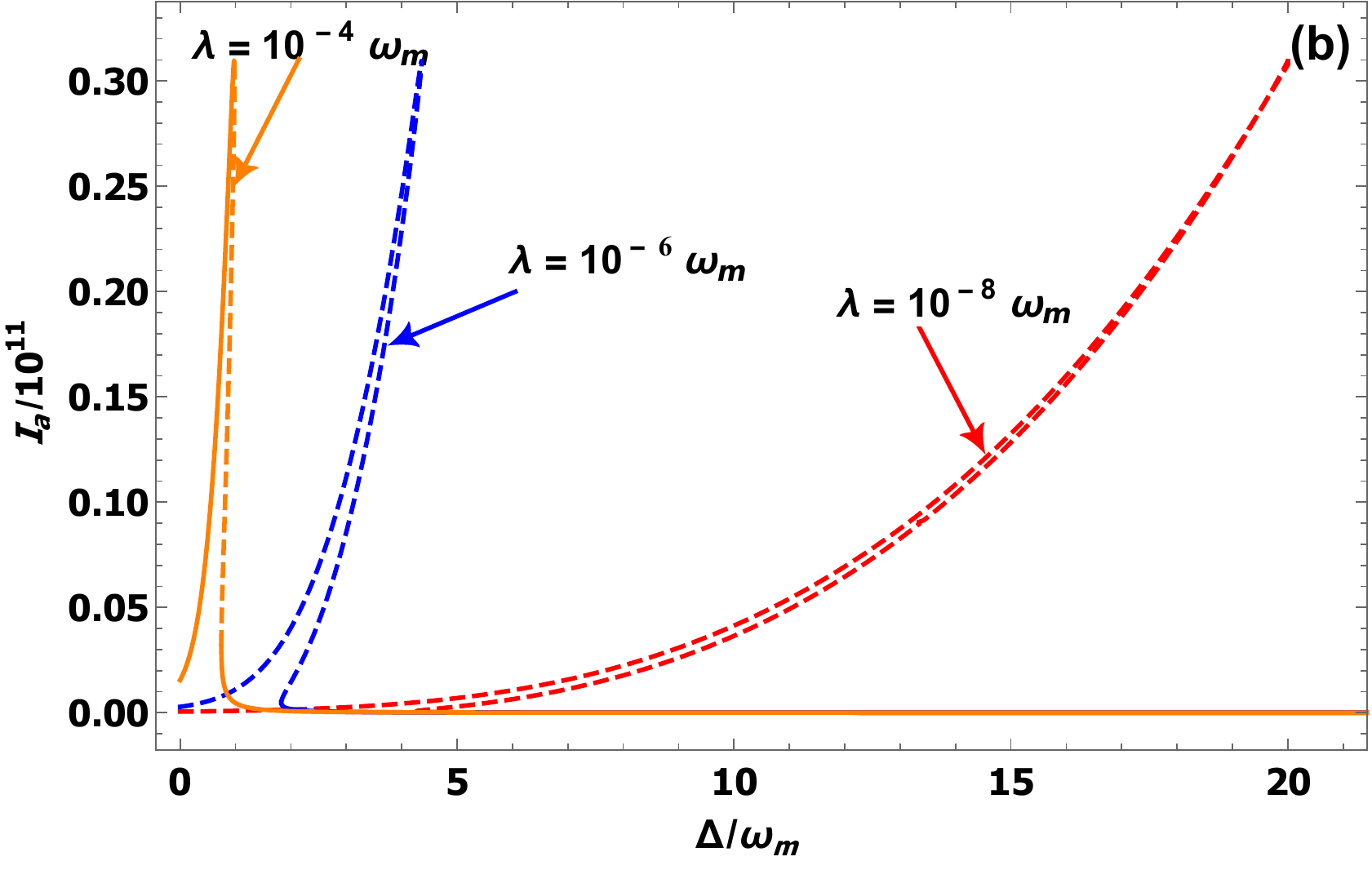}
\caption{\label{fig3}(a) Steady-state mechanical oscillation amplitude and (b) steady-state intracavity intensity versus the normalized bare detuning $\Delta/\omega_m$ in the presence of the gain nonlinearity for  $\omega_m/2\pi= 2\text{MHz}$, $\kappa_c=0.2\omega_m$, $G_0=0.3\kappa_c$, $\theta=\pi/8$, and $P_{in}=3\text{mW}$. Other parameters are the same as those
in Fig.~\ref{fig2}.}%
\end{figure}

\begin{figure}[h!]
 \centering
   \includegraphics[scale=0.37]{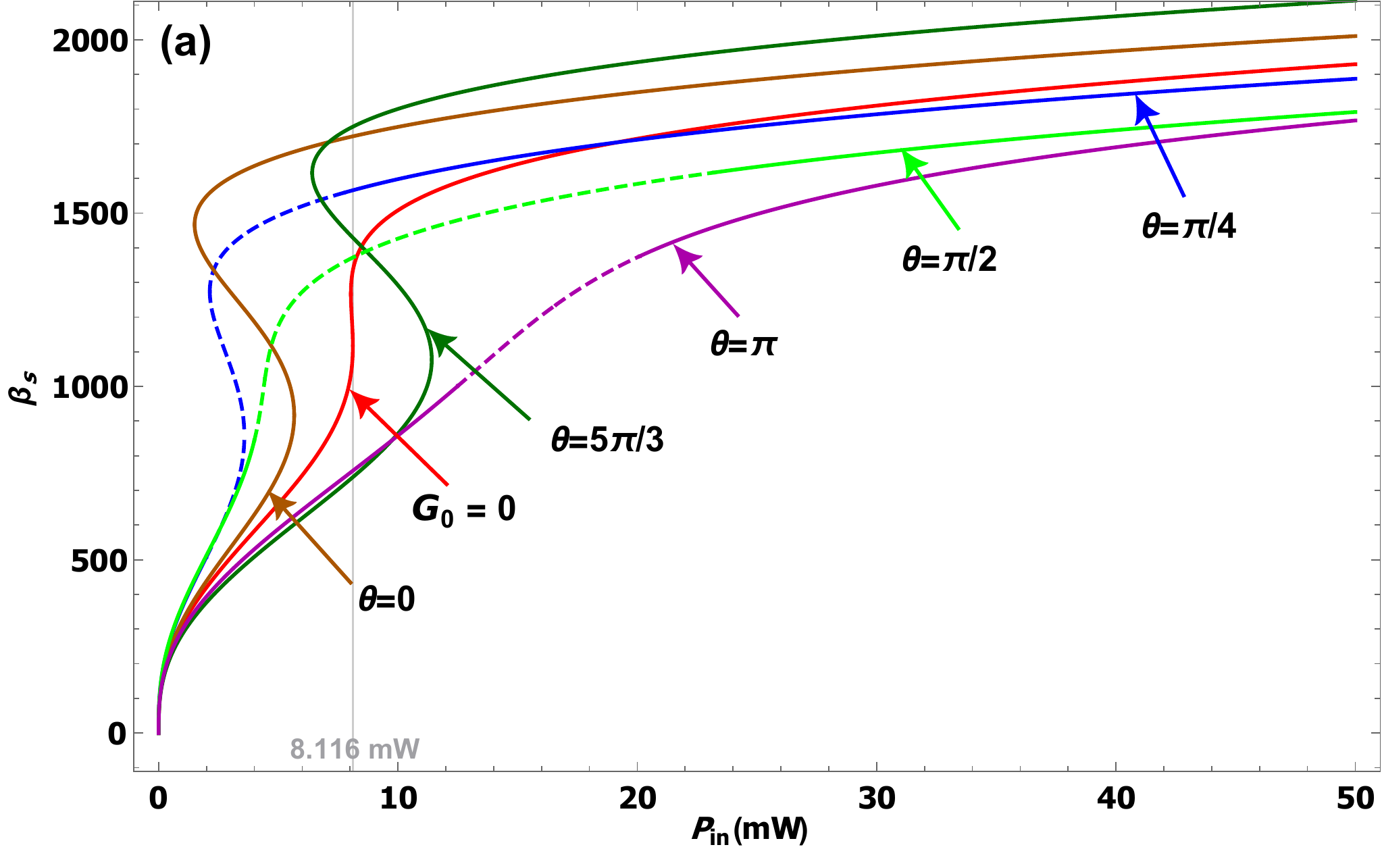} \par
  \includegraphics[scale=0.37]{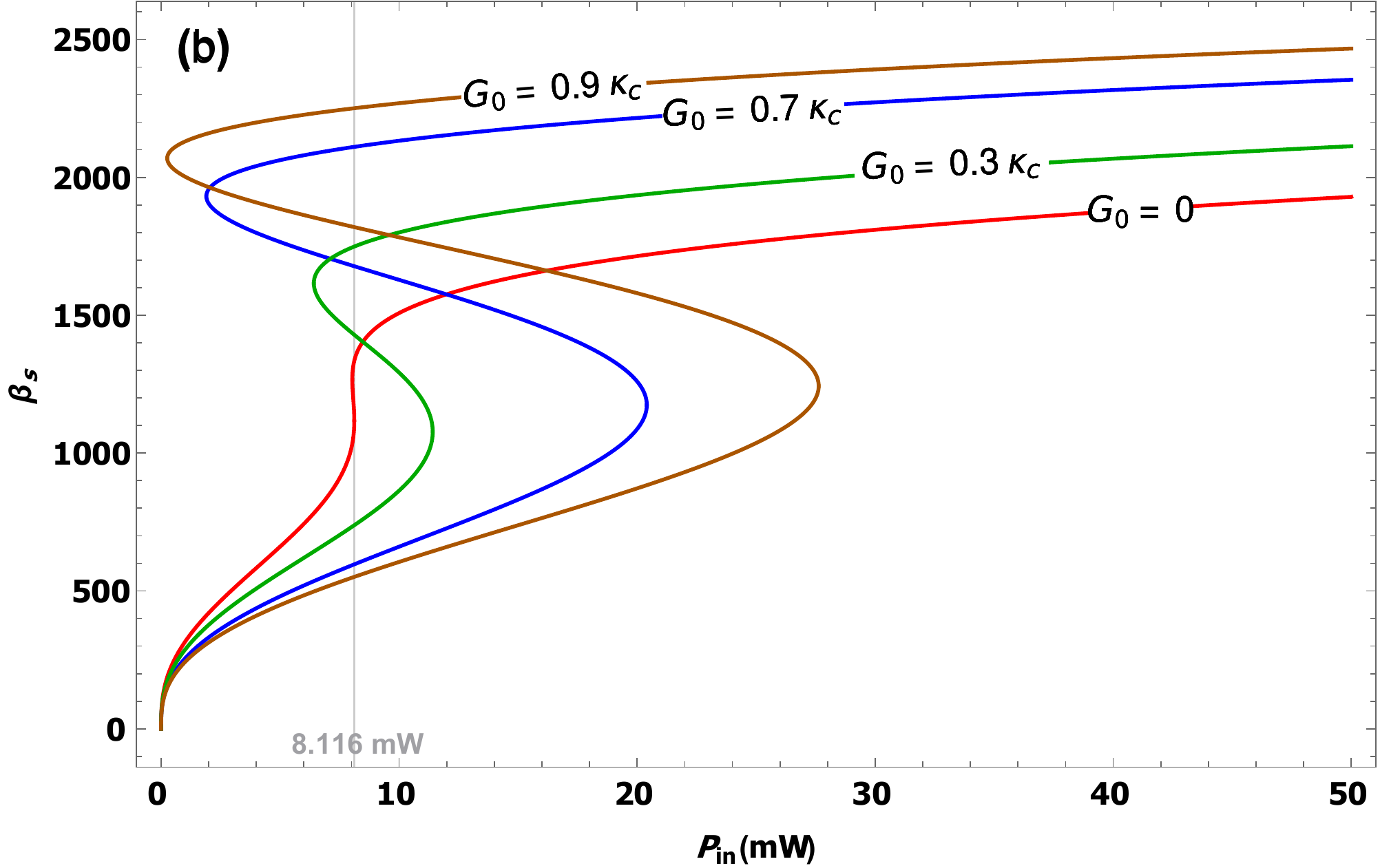}
\caption{ \label{fig4}Steady-state mechanical oscillation amplitude versus  the input power $P_{in}$ at $\Delta \approx \Delta^{crit}=0.7998 \omega_m$ with $\lambda=10^{-4}\omega_m$ for (a) $G_0=0.3\kappa_c$ and different values of the phase $\theta$, and for (b) $\theta=5\pi/3$ and different values of the  nonlinear gain $G_0$. Other parameters are the same as those
in \cref{fig3}. The critical value of the input laser power for the appearance of the mechanical multistability in the absence of the gain nonlinearity ($G_0=0$) has also been shown.}%
 \end{figure}

Now, we consider the general case with nonzero (and positive) Duffing nonlinearity strength $\lambda$.  In such a case, as can be seen in Eqs.~(\ref{criteqs1}-\ref{criteqs3}), the non-zero value of the Duffing nonlinearity results in the appearance of the higher powers of $\bar{k}$ which have important contributions when $\bar{k}$ is sufficiently large. Therefore, the Duffing nonlinearity together with  $\bar{k}\gg 1$ lead  to an increase in the critical  values of the system.
In order to investigate the effect of the Duffing nonlinearity on the steady-state intracavity intensity, $I_a$,  and the mechanical oscillation amplitude, $\beta_s$,  in Fig.~\ref{fig2} we have plotted these two quantities versus the normalized bare cavity detuning $\Delta/\omega_m$ for a bare cavity ($G_0=0$). As is seen, in the presence of the Duffing nonlinearity the mechanical oscillation amplitude is suppressed significantly (Fig.~\ref{fig2}(a)), while the resonance frequency of the cavity is only a little bit shifted to the lower values (Fig.~\ref{fig2}(b)). 
Figure \ref{fig3} illustrates the impact of the  Duffing anharmonicity on the behaviors of the steady-state values $I_a$ and $\beta_s$ as functions of the bare detuning $\Delta/\omega_m$ when $G_0=0.3\kappa_c$ and $\theta=\pi/8$. As can be seen, the mechanical anharmonicity causes the width of the multi- solution region to be significantly reduced, while the effect of the nonlinear gain medium manifests itself in pushing the steady-state values $I_a$ and $\beta_s$ to the multi- solution region. Also, it is clear that with increasing the Duffing parameter $\lambda$ the mechanical oscillation amplitude is suppressed (Fig. \ref{fig3}(a)), while the maximum available value of the intracavity intensity does not change (Fig. \ref{fig3}(b)).
However, the nonlinear gain, $G_0$, and the phase of the field driving the OPA, $\theta$, can be used to control and manipulate the optical and mechanical bistability behaviors in the system. Figure \ref{fig4}(a) shows the steady-state mechanical oscillation amplitude versus the input power $P_{in}$ for different values of the phase $\theta$ with $G_0=0.3\kappa_c$ and $\lambda=10^{-4}\omega_m$. In \cref{fig4}(b) the steady-state mechanical oscillation amplitude is plotted as a function of the input laser power $P_{in}$ and nonlinear gain $G_0$ with $\theta=5\pi/3$ and $\lambda=10^{-4}\omega_m$. In  these figures we choose the bare detuning $\Delta$ to be equal to its critical value, $\Delta=0.7998 \omega_m$. In addition, the threshold value of the input laser power to observe multi- solution region in the absence of the OPA ($8.116 \text{mW}$) has been shown.
As can be seen, in the presence of the OPA  the critical value of the input laser power to observe bistable behavior as well as the width of  the multistability region denpends on the values of $\theta$ and $G_0$. In addition, by adjusting $\theta$ and $G_0$, the OPA can be used to control the suppression of the steady-state mechanical oscillation amplitude caused by the Duffing anharmonicity.

 To sum up this subsection, we have found that the presence of the OPA and Duffing-like anharmonicity of the mechanical mode in an optomechanical cavity can greatly alter the critical quantities of the system to observe optical and mechanical multistabilities. The Duffing-type of mechanical nonlinearity in the system causes the steady-state mechanical oscillation amplitude to be suppressed but does not affect the steady-state mean number of the intracavity photons. Therefore, the conventional linearization procedure for the cavity-field operators can be safely done but the validity of the linearization approximation for the operators of the mechanical mode imposes a new assumption on the system parameters which will be explained in the following. Moreover, the nonlinear gain medium can be used to control and manipulate the multistable behavior of the system as well as the magnitude of the steady-state optical and mechanical amplitudes. This is an advantage to produce strong steady-state mechanical squeezing in  the system under consideration with a fixed input laser power.
\subsection{Small fluctuations dynamics}
Having discussed the mean-field solutions, we now proceed to examine the fluctuations dynamics. The linearized quantum Langevin equations for the fluctuation operators read as
\vskip -0.85cm
\begin{subequations}\label{linqle}
\begin{align}
&\delta \dot{\hat{a}}=-i\Delta' \delta \hat{a }+ i \frac{\bar{G}}{2}\,(\delta \hat{b}+\delta \hat{b}^\dagger)\nonumber\\&\hspace*{2cm}+2G_0 e^{i\theta}\delta \hat{a}^\dagger  - \kappa_c\, \delta \hat{a}+\sqrt{2\kappa_c}\,\hat{a}_{in}\label{vara},\\&\delta \dot{\hat{b}}=-i(\omega_m+2\Lambda) \delta\hat{b} + i \frac{\bar{G}}{2}\,(\delta \hat{a}+ \delta \hat{a}^\dagger)\nonumber\\&\hspace*{2cm}-2i \Lambda \delta \hat{b}^\dagger - \gamma_m\,\delta \hat{b}\,+\sqrt{2\gamma_m} \,\hat{b}_{in},\label{varb}
\end{align}
\end{subequations}\vskip -0.6cm
\noindent where $\Lambda=3\lambda(1+4 \beta_s^2)$ and $\bar{G}=2g \alpha_s$ with $\alpha_s=\langle \hat{a}\rangle_s$ are the enhanced Duffing parameter and the coherent intracavity-field-enhanced optomechanical coupling strength, respectively. It is worth to mention that in the linearization of  \cref{beq} we have conidered $\beta_s\gg1$ and $\lambda \beta_s\ll \Lambda,\bar{G}$ so that the Duffing nonlinear term can be approximately written as
\begin{align}
&\hspace*{-0.2cm} \lambda (\hat{b}+\hat{b}^\dagger)^3=\lambda\Big[8\beta_s^3 \,+\, 6\beta_s\,+\,3(1+4 \beta_s^2)(\delta \hat{b}+\delta \hat{b}^\dagger)\,\nonumber\\&\hspace*{2cm}+: (\delta \hat{b}+\delta \hat{b}^\dagger)^3:\,+\, 6\beta_s :(\delta \hat{b}+\delta \hat{b}^\dagger)^2: \Big]\nonumber\\&\hspace*{1.2cm}\approx\lambda\Big[8\beta_s^3 \,+\, 6\beta_s\,+\,3(1+4 \beta_s^2)(\delta \hat{b}+\delta \hat{b}^\dagger) \Big].
\end{align}
Moreover, as already shown in the previous subsection, the presence of the Duffing nonlinearity results in a lowering of the mechanical oscillation amplitude. Therefore, the fulfillment of the conditions $\beta_s\gg1$ and $\lambda \beta_s\ll \Lambda,\bar{G}$ must be carefully checked for each set of parameters used in numerical calculations.

Equation (\ref{varb}) shows that  in the linear approximation  the Duffing nonlinearity leads to the frequency shift  as well as parametric amplification of the mechanical mode. For a bare optomechanical cavity ($G_0=0$)  it has been shown theoretically \cite{o45} that the joint effect of this nonlinearity-induced parametric amplification and cavity cooling can result in a strong mechanical squeezing which is robust against thermal fluctuations of the mechanical mode. To simplify the subsequent calculations, we apply the unitary transformation $\hat{S}(r)=\text{exp}[\frac{r}{2}(\delta \hat{b}^2-\delta \hat{b}^{\dagger^2})]$, which is the single-mode squeezing operator with the squeezing parameter $r=\frac{1}{4}\text{ln}[1+\frac{4\Lambda}{\omega_m}]$, to Eqs.~(\ref{vara}) and (\ref{varb}). Under this transformation,
\begin{subequations}\begin{align}
&\hat{S}(r)\delta \hat{ b} \hat{ S}^\dagger(r)=\cosh(r) \delta \hat{ b} -\,\sinh(r) \delta \hat{ b}^\dagger,\\&\hat{S}(r)\delta \hat{ b}^\dagger \hat{ S}^\dagger(r)=\cosh(r) \delta \hat{ b} ^\dagger-\,\sinh(r) \delta \hat{ b},
\end{align}\end{subequations}
and thus the linearized quantum Langevin equations for the fluctuations are transformed to
\begin{subequations}\label{rlqle}
\begin{align}
&\dot{ \delta \hat{ a}}=-i\Delta' \delta\hat{a} + i\, \frac{\bar{G'}}{2}\,(\delta \hat{ b}+\delta \hat{ b}^\dagger)+2G_0 e^{i\theta}\delta \hat{ a}^\dagger\nonumber\\&\hspace*{4cm}  - \kappa_c\, \delta \hat{a}+\sqrt{2\kappa_c}\,\hat{a}_{in},\\& \dot{ \delta \hat{ b}}=-i\,\Omega_m \delta \hat{ b} + i \,\frac{\bar{G'}}{2}\,(\delta \hat{ a}+ \delta \hat{a}^\dagger)\nonumber\\  &\hspace*{4cm} - \gamma_m \,\delta \hat{ b}\,+\sqrt{2\gamma_m} \,\tilde{b}_{in},
\end{align}
\end{subequations}
where $\Omega_m=e^{2r}\omega_m $ is the transformed effective mechanical frequency, $\bar{G'}=e^{-r}\,\bar{G}$ is the transformed effective optomechanical coupling, and $\hat{\tilde{b}}_{in}=\cosh(r) \,\hat{b}_{in}\,+\,\sinh(r) \,\hat{b}_{in}^\dagger$ is the input thermal noise of the mechanical mode in the  new rotating frame.

By defining the cavity-field quadratures fluctuations as $\delta \hat{x}=\frac{\delta \hat{a} +\delta \hat{a}^\dagger}{\sqrt{2}}$ and $\delta\hat{y}=\frac{\delta \hat{a} -\delta \hat{a}^\dagger}{\sqrt{2}i}$, and also the mechanical quadratures fluctuations   as $\delta\hat{q}=\frac{\delta \hat{b} +\delta \hat{b}^\dagger}{\sqrt{2}}$ and $\delta\hat{p}=\frac{\delta \hat{b} -\delta \hat{b}^\dagger}{\sqrt{2}i}$ the equations of motion (\ref{vara}) and (\ref{varb}) can be written in the compact matrix form
\begin{equation}\label{comqle}
\delta\dot{\hat{u}}(t)=\mathbf{A}.\delta\hat{u}(t)\,+\, \delta\hat{n}(t),
\end{equation}
where $\delta\hat{u}(t)=(\delta\hat{x},\,\delta\hat{y},\,\delta\hat{q},\,\delta\hat{p})^T$ is the vector of continuous-variable fluctuation operators and 
$\delta\hat{n}(t)=(\sqrt{2\kappa_c}\,\delta\hat{x}_{in},\,\sqrt{2\kappa_c}\,\delta\hat{y}_{in},\,\sqrt{2\gamma_m}\,\delta\hat{q}_{in},\,\sqrt{2\gamma_m}\,\delta\hat{p}_{in})^T$ is the corresponding vector of noises in which $\delta \hat{x}_{in}=\frac{\hat{a}_{in} + \hat{a}^\dagger_{in}}{\sqrt{2}}$ and $\delta\hat{y}_{in}=\frac{\hat{a}_{in} - \hat{a}^\dagger_{in}}{\sqrt{2}i}$ denote the input noise quadratures of the optical field, and $\delta\hat{q}_{in}$\begin{scriptsize}$=\dfrac{\hat{\tilde{b}}_{in} +\hat{\tilde{b}}_{in}^\dagger}{\sqrt{2}}$\end{scriptsize} and $\delta\hat{p}_{in}$\begin{scriptsize} $=\dfrac{\hat{\tilde{b}}_{in} -\hat{\tilde{b}}^\dagger_{in}}{\sqrt{2}i}$\end{scriptsize} refer to the input noise quadratures of the mechanical mode of the moving mirror. Moreover, the 4 × 4 matrix $\mathbf{A}$ is the drift matrix given by
\begin{equation}{\label{drift}}
\mathbf{A}=\begin{pmatrix}
-(\kappa_c -\kappa_p)&\Delta' +\Delta_p & 0 & 0\cr -(\Delta' -\Delta_p)& -(\kappa_c + \kappa_p) &\bar{G'}& 0 \cr 0&0&-\gamma_m & \Omega_m \cr \bar{G'}&0&-\Omega_m  &-\gamma_m
\end{pmatrix},
\end{equation}
in which $\Delta_p=2G_0\,\sin\theta$ and $\kappa_p=2G_0\,\cos\theta$.

The stationary properties of the system fluctuations can be explored by considering the steady-state condition governed by \cref{comqle}. The steady state associated with \cref{comqle} is reached when the system is stable, which occurs if and only if all the eigenvalues of the drift matrix $\mathbf{A}$ have negative real parts. These stability conditions can be obtained by using the Routh–Hurwitz criterion \cite{o64}, which results in the following independent conditions on the system parameters:
\begin{subequations}\label{RH}
\begin{align}
&\hspace{-0.1cm}s_1=\gamma_m\{(2\kappa_c +\gamma_m)^2\,+\,e^{4r}\omega_m^2\}\nonumber\\&\hspace*{2.5cm}+\kappa_c\Big(\Delta{'}^2+\kappa_c^2-4G_0^2\Big)>0,\label{ss1}\\& \hspace{-0.1cm}s_2=(\Delta{'}^2+\kappa_c^2-4G_0^2)(e^{4r}\omega_m^2\,+\gamma_m ^2)\,\nonumber\\&\hspace*{3cm}-\, \omega_m(\Delta'+\Delta_p)\bar{G}^2>0, \label{dc}\\ &\hspace{-0.1cm} s_3=\gamma_m  \kappa_c\Big \{(\Delta{'}^2+\kappa_c^2-4 G_0 ^2- e^{4r}\omega_m^2+\gamma_m(\gamma_m+2\kappa_c))^2\,\nonumber\\&\hspace*{3cm}+\,4\gamma_m \kappa_c (\gamma_m +\kappa_c)^2 e^{4r}\omega_m^2\Big\}\nonumber\\&\hspace*{1.5cm}+(\gamma_m +\kappa_c)^2
\omega_m (\Delta'\,+\,\Delta_p)\bar{G}^2>0,\label{Ss3}
\end{align}
\end{subequations}
The first condition is satisfied if we require that $(\Delta{'}^2+\kappa_c^2-4G_0^2)>0$  which is always satisfied for a bare cavity ($G_0=0$) and gives the threshold condition for parametric oscillation. The violation of the second condition ($s_2<0$) leads to the instability in the region $(\Delta'+\Delta_p)>0$. Also, for a bare optomechanical cavity this condition results in a bistability in the system. Since the expression in the  brackets in Eq.~(\ref{Ss3}) is always positive, the violation of the third condition ($s_3<0$) implies that we have $(\Delta'+\Delta_p)<0$. For a bare cavity this condition causes the instability in the domain of the blue- detuned laser. In the following, we assume that $(\Delta'+\Delta_p)>0$, so the stability of the system is determined only by $s_1$ and $s_2$.

\section{ \label{sec:3th}Mechanical ground-state cooling and quadrature squeezing }
In this section we are going to investigate how the mechanical anharmonicity and the gain nonlinearity affect  the ground-state cooling as well as the quadrature squeezing of the movable mirror. For this purpose, we need to obtain the mean square of fluctuations of the mechanical quadratures. To this end, we write Eq.~(\ref{comqle}) in the Fourier space by using
\begin{equation}
\delta\hat{O}(t)=\frac{1}{2\pi}\int_{-\infty}^{\infty}\,d\,\omega \,e^{-i\omega\,t}\delta\hat{\tilde{O}}(\omega),
\end{equation}
and solve it in the frequency domain to obtain the following expressions for the quantum fluctuations of the movable mirror in the transformed frame
\begin{subequations}
\begin{align}
&\delta\hat{\tilde{q}}(\omega)=A_1(\omega)\delta\hat{\tilde{x}}_{in}(\omega)+A_2(\omega)\delta\hat{\tilde{y}}_{in}(\omega)\nonumber\\&\hspace*{2cm}+A_3(\omega)\delta\hat{\tilde{q}}_{in}(\omega)+A_4(\omega)\delta\hat{\tilde{p}}_{in}(\omega)\label{a},\\&\delta\hat{\tilde{p}}(\omega)=B_1(\omega)\delta\hat{\tilde{x}}_{in}(\omega)+B_2(\omega)\delta\hat{\tilde{y}}_{in}(\omega)\nonumber\\&\hspace*{2cm}+B_3(\omega)\delta\hat{\tilde{q}}_{in}(\omega)+B_4(\omega)\delta\hat{\tilde{p}}_{in}(\omega)\label{b},
\end{align}\end{subequations}
where 
\begin{subequations}
\begin{align}
&A_1(\omega)=\frac{\sqrt{2\kappa_c}}{d(\omega)}\,e^{r}\bar{G}\omega_m(\kappa_c-i \omega +\Delta_p),\\&A_2(\omega)=\frac{\sqrt{2\kappa_c}}{d(\omega)} e^{r}\bar{G}\omega_m(\Delta'+\Delta_p),\\&A_3(\omega)=\frac{\sqrt{2\gamma_m}}{d(\omega)}(\gamma_m-i\omega)(\Delta{'}^2+(\kappa_c-i\omega)^2-4G_0^2),\\&A_4(\omega)=\frac{e^{2r}\omega_m}{(\gamma_m-i\omega)}A_3(\omega),\\&B_1(\omega)=\frac{1}{d(\omega)}\frac{\gamma_m-i\omega}{e^{2r}\omega_m}A_1,\\&B_2(\omega)=\frac{\sqrt{2\kappa_c}}{d(\omega)}\,e^{-r}\bar{G}(\gamma_m-i\omega)(\Delta'+\Delta_p),\\&B_3(\omega)=\frac{\sqrt{2\gamma_m}}{d(\omega)}\Big(e^{-2r}\bar{G}^2(\Delta'+\Delta_p)\nonumber\\&\hspace*{1cm}- e^{2r}\omega_m(\Delta{'}^2+(\kappa_c-i\omega)^2-4G_0^2)\Big),\\&B_4(\omega)=A_3(\omega),
\end{align}
\end{subequations}
with
\begin{subequations}\begin{align}
&\hspace*{-0.3cm}d(\omega)=(\Delta{'}^2+(\kappa_c-i\omega)^2-4G_0^2)\chi^{-1}_m(\omega)\nonumber\\&\hspace*{4cm}-\omega_m\bar{G}^2(\Delta'+\Delta_p),\\&\hspace*{-0.3cm}
\chi^{-1}_m(\omega) =\gamma_m^2-2i \gamma_m \omega-\omega^2+\,e^{4r}\omega_m^2.
\end{align}
\end{subequations} 
Furthermore, the input noise quadratures of the optical and mechanical modes satisfy the following correlation functions
in the frequency domain:
\begin{subequations}
\begin{align}
&\hspace*{-0.4cm}\langle \delta\hat{\tilde{x}}_{in}(\omega)\delta\hat{\tilde{x}}_{in}(\Omega)\rangle =\langle \delta\hat{\tilde{y}}_{in}(\omega)\delta\hat{\tilde{y}}_{in}(\Omega)\rangle = \pi \delta(\omega+\Omega)\\&\hspace*{-0.4cm}\langle \delta\hat{\tilde{x}}_{in}(\omega)\delta\hat{\tilde{y}}_{in}(\Omega)\rangle= i \,\pi \delta(\omega+\Omega)\\&\hspace*{-0.4cm}\langle \delta\hat{\tilde{q}}_{in}(\omega)\delta\hat{\tilde{q}}_{in}(\Omega)\rangle = \pi\,e^{2r}(1+2\bar{n}_m) \delta(\omega+\Omega)\\&\hspace*{-0.4cm}\langle \delta\hat{\tilde{p}}_{in}(\omega)\delta\hat{\tilde{p}}_{in}(\Omega)\rangle= \pi\,e^{-2r}(1+2\bar{n}_m)  \delta(\omega+\Omega)\\&\hspace*{-0.4cm}\langle \delta\hat{\tilde{q}}_{in}(\omega)\delta\hat{\tilde{p}}_{in}(\Omega)\rangle= i \,\pi \delta(\omega+\Omega)
\end{align}
\end{subequations}
In each of Eqs.~(\ref{a}) and (\ref{b}), the first two terms arise from the radiation pressure while the other two terms originate from the mechanical thermal noise. The mean square of fluctuations are determined by 
\begin{equation}\label{spov}
\langle (\delta \hat{O}(t))^2\rangle=\frac{1}{2\pi}\int_{-\infty}^{\infty}\,d\,\omega \, S_O(\omega),\quad (O=q,p),
\end{equation}
in which $S_O(\omega)$ is the symmetrized spectrum of fluctuation in operator $\hat{O}$ and is defined by~\cite{o65}
\begin{align}\label{spo}
&S_O(\omega)=\frac{1}{4\pi}\int_{-\infty}^{\infty}\,d\,\Omega \, e^{-i(\omega+\Omega)t}\langle\delta \hat{O}(\omega)\delta \hat{O}(\Omega)\nonumber\\&\hspace*{3cm}+\delta \hat{O}(\Omega)\delta \hat{O}(\omega) \rangle,\quad (O=q,p).
\end{align}

To examine the mechanical ground-state cooling, we consider the effective steady-state mean phonon number of the mechanical mode in the original (untransformed) frame which is defined as $n_{\text{eff}}=[\langle( \delta\hat{q})^2\rangle+\langle( \delta\hat{p})^2\rangle-1]/2$ (corresponding to an effective mode temperature $T_{\text{eff}}=\hbar \omega_m/[k_{B} \,\text{ln}(1+1/n_{\text{eff}})]$). The ground-state cooling is achieved whenever $n_{\text{eff}}\simeq 0$ which occurs if $\langle( \delta\hat{q})^2\rangle\simeq\langle( \delta\hat{p})^2\rangle\simeq1/2$ in steady state.

To explore the impacts of the mechanical and optical nonlinearities on the squeezing of the position and momentum quadratures of the movable mirror we consider the degree of squeezing which in the dB (decibel) unit can be calculated by~\cite{o34}~\begin{small}$D_O= -10 \,\text{log}_{10}\frac{\langle( \delta\hat{O})^2\rangle}{\,\,\langle( \delta\hat{O})^2\rangle_{vac}},\,(\hat{O}=\hat{q},\hat{p})$\end{small} with \begin{small}$\langle( \delta\hat{O})^2\rangle_{vac}=1/2$\end{small} as the quantum-vacuum fluctuation. Whenever, \begin{small}$D_O>0$\end{small}, the corresponding mechanical quadrature is a squeezed one.

The steady-state variances of the mechanical quadrature  fluctuations in the original (untransformed) frame can be obtained as \begin{small}$\langle( \delta\hat{q})^2\rangle=e^{-2r} (n'_{\text{eff}}+\frac{1}{2})$\end{small} and \begin{small}$\langle( \delta\hat{p})^2\rangle=e^{2r} (n'_{\text{eff}}+\frac{1}{2})$\end{small} where \begin{small}$n'_{\text{eff}}$\end{small} is the  transformed steady-state phonon number. To get a strong mechanical squeezing  $n'_{\text{eff}}$ must be small as far as possible, which means that the mechanical motion in the rotated frame must be cooled down. The best cooling in the transformed system and so the strongest mechanical squeezing is occured at the optimal detuning \begin{small}$\Delta'\approx \Omega_m$\end{small}~\cite{o45}. Also, it is worth to stress that to establish the validity of the linearization procedure in the Duffing term, we must solve Eqs. (\ref{as}) and (\ref{bs}) numerically and determine the range of the desired variable (bare detuning or Duffing parameter) over which the assumption \begin{small}$\beta_s\gg 1$\end{small} holds (in fact, we consider \begin{small}$\beta_s\geq 40$\end{small}).  

Although the explicit forms of the momentum and position variances are too complicated to be treated analytically we can obtain the following approximate expressions for them in the limit of large mechanical quality factor (\begin{small}$Q_m\gg 1$\end{small}) and low-temperature environment (\begin{small}$\kappa_c\gg \gamma_m \bar{n}_m$\end{small}) in the original (untransformed) frame:
\begin{small}\begin{subequations}\label{aqp}
\begin{align}
&\hspace*{-0.5cm}\langle( \delta\hat{q})^2\rangle =\frac{\omega_m}{4(\Delta'+\Delta_p)}+\frac{e^{-4r}\{(\Delta'+\Delta_p)^2+(\kappa_c+\kappa_p)^2\}}{4\eta\,\omega_m\,(\Delta'+\Delta_p)},\label{aqp1}\\&\hspace*{-0.5cm}\langle( \delta\hat{p})^2\rangle = \frac{e^{2r}}{2}+\frac{(\Delta'+\Delta_p-\omega_m\,e^{2r})^2+(\kappa_c+\kappa_p)^2}{4\,\omega_m\,(\Delta'+\Delta_p)},\label{aqp2}
\end{align}
\end{subequations}
\end{small}
\noindent\hspace*{-0.3cm} where $\eta$ is the so-called ``bistability parameter''~\cite{o65} and is defined by
\begin{align}\label{eta}
&\hspace*{-0.4cm}\eta= 1-\frac{\bar{G}^2(\Delta'+\Delta_p)}{\omega_m\,e^{4r}(\Delta'^2+\kappa_c^2-4G_0^2)}.
\end{align}

The first point that can be inferred from Eqs.~(\ref{aqp1}) and (\ref{aqp2}) is that for $\Delta'+\Delta_p>0$ there is no squeezing in the momentum fluctuation of the movable mirror. It is easy to verify this claim by setting the value of $\Delta'+\Delta_p$ to $\omega_m\,e^{2r}$ and the value of $\kappa_c$ to $-\kappa_p$ (for $\pi/2<\theta<3\pi/2$). With these choices and for $\eta\simeq 1$, Eqs.~(\ref{aqp1}) and (\ref{aqp2}) are reduced to $\langle(\delta\hat{q})^2\rangle=\frac{e^{-2r}}{2}$ and $\langle( \delta\hat{p})^2\rangle=\frac{e^{2r}}{2}$, respectively. Thus for $\Delta' +\Delta_p>0$, in the absence of the Duffing anharmonicity ($r=0$) there is no squeezing in the mechanical quadratures fluctuations. Also, since $r$ is a function of the input laser power $P_{in}$ and properties of the optical parametric nonlinearity, then the position  quadrature squeezing created by the Duffing nonlinearity can be controlled by $P_{in}$, $G_0$, and $\theta$.

Another important point is the presence of the factor $e^{-4r}/\eta$ in Eq.~(\ref{aqp1}). According to Eq.~(\ref{eta}), with increasing the input laser power the value of $\eta$ differs from 1. For $\eta<1$, the existence of factor $e^{-4r}/\eta$ leads to a significant difference between the minimum available number of phonons in our system and in its harmonic counterpart.

For $r=0$, Eqs.~(\ref{aqp}) and (\ref{eta}) are the same as those reported in Ref.~\cite{o32}. As the authors have discussed in this reference, for $\eta\simeq 1$ the steady-state mean phonon number tends to it minimum value at \begin{small}$\Delta'_{opt}=-\Delta_p+\sqrt{\omega_m^2+(\kappa_c+\kappa_p)^2}$\end{small}. To get a preliminary insight into the effect of the squeezing parameter $r$ (and so the Duffing anharmonicity parameter) on this minimum attainable value, by assuming\begin{small} $\Delta'=\Delta'_{opt}$\end{small} as well as using exponential expansion for $e^{2r}$ and $e^{-4r}$ in Eq.~(\ref{aqp}), we can acquire the steady-state mean phonon number $n_{\text{eff}}$ for $\eta\simeq 1$ as follows
\begin{align}\label{nefo1}
&\hspace*{-0.4cm}n_{\text{eff}}=- \frac{1}{2}+\frac{1}{2}\sqrt{1+(\frac{\kappa_c+\kappa_p}{\omega_m})^2}\Big( 1+4r^2+\frac{16}{3}r^4+\cdots\Big)\nonumber\\&\,-\,(\frac{\kappa_c+\kappa_p}{\omega_m})^2\frac{1}{\sqrt{1+(\frac{\kappa_c+\kappa_p}{\omega_m})^2}}\Big( r+\frac{8}{3}r^3+\frac{32}{15}r^5+\cdots\Big).
\end{align}
If $r\ll 1$, neglecting the terms higher than the second order of $r$ leads to an expression for $n_{\text{eff}}$ which can be minimized with respect to the squeezing parameter $r$. In this way, the optimum value of $r$ reads
\begin{align}\label{ropt}
&r_{opt}=\frac{1}{4}\frac{(\kappa_c+\kappa_p)^2}{(\kappa_c+\kappa_p)^2+\omega_m^2},
\end{align}
and the minimum attainable value of $n_{\text{eff}}$ can be obtained as follows
\begin{align}\label{nef2}
&\hspace*{-1cm}n_{\text{eff},\text{min}}=\frac{1}{2}\Big(-1+\sqrt{1+\frac{(\kappa_c+\kappa_p)^2}{\omega_m^2}}\Big)\nonumber\\&\,-\,\frac{1}{8}\Big(\frac{(\kappa_c+\kappa_p)^4}{\omega_m^4}\frac{1}{(1+\frac{(\kappa_c+\kappa_p)^2}{\omega_m^2})^{3/2}}\Big).
\end{align}
From the above equation, it is clear that the presence of the Duffing anharmonicity can lead to a decrease in the minimum attainable number of phonons, which is dependent on $(\kappa_c+\kappa_p)/\omega_m$. In addition, according to Eq.~(\ref{ropt}) the optimum value of the squeezing parameter $r_{opt}$, which depends on the ratio $(\kappa_c+\kappa_p)/\omega_m$, is controllable by the properties of the nolinear gain medium. 

In the general case, the effect of the temperature, $\eta\neq 1$ as well as arbitrary values of $r$ must be taken into account, so we can not obtain the explicit analytical relation for the minimum value of the steady-state mean phonon number and inevitably we must examine the problem numerically. To investigate the effect of  the mechanical anharmonicity on the effective mode temperature and the degree of the mechanical squeezing, we first examine its impact on the mean square of the momentum and displacement quadratures fluctuations of the movable mirror. 
\begin{figure}[h!]
 \centering
   \includegraphics[scale=0.48]{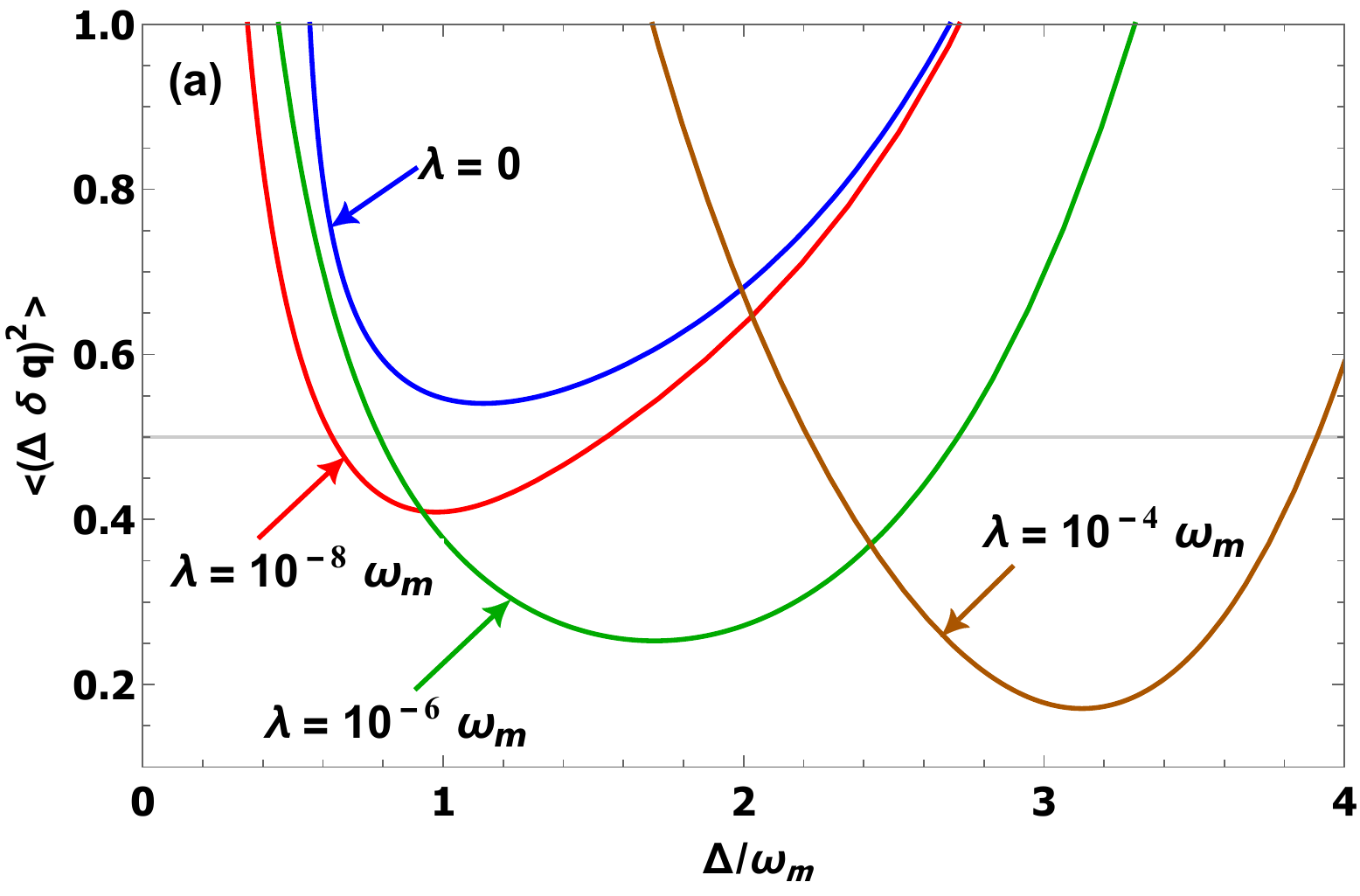} \par
  \includegraphics[scale=0.48]{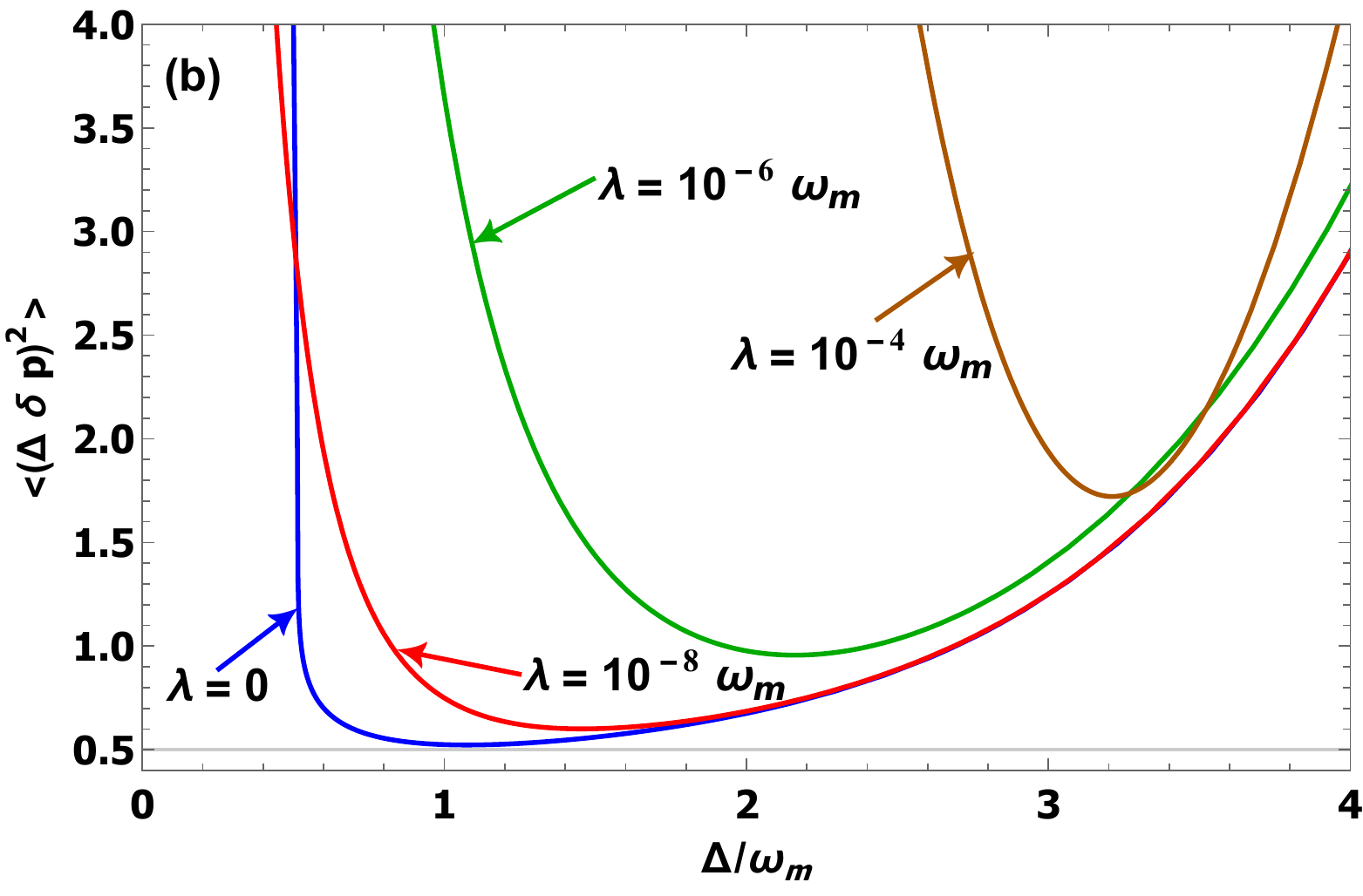}
\caption{ \label{fig5} The steady-state variances of (a) the displacement quadrature fluctuation and (b) the momentum quadrature fluctuation of the moving mirror versus the normalized bare detuning $\Delta/\omega_m$  for a bare cavity ($G_0=0$) with $\omega_m/2\pi= 10\text{MHz}$, $Q_m=10^{6}$, $\kappa_c=0.3\omega_m$, $\lambda_L=1064\,\text{nm}$, $P_{in}=3\text{mW}$ and $T=25\,\text{mK}$. Other parameters are the same as those in \cref{fig4}.}%
\end{figure}
In Fig. \ref{fig5} we have plotted the steady-state mean square of these quadratures against the normalized bare detuning $\Delta/\omega_m$  for a bare cavity and for different values of the mechanical nonlinearity  parameter $\lambda$. As can be seen from \cref{fig5}(a), the mechanical Duffing anharmonicity can lead to the position squeezing of the mirror ($\langle( \delta\hat{q})^2\rangle<1/2$); the larger the Duffing anharmonicity parameter is, the stronger the position quadrature squeezing is. In contrast, as is shown in \cref{fig5}(b), there is no squeezing in the momentum fluctuation of the movable mirror and the momentum variance increases as the Duffing parameter increases.
\begin{figure}[h!!]
 \centering
   \includegraphics[scale=0.48]{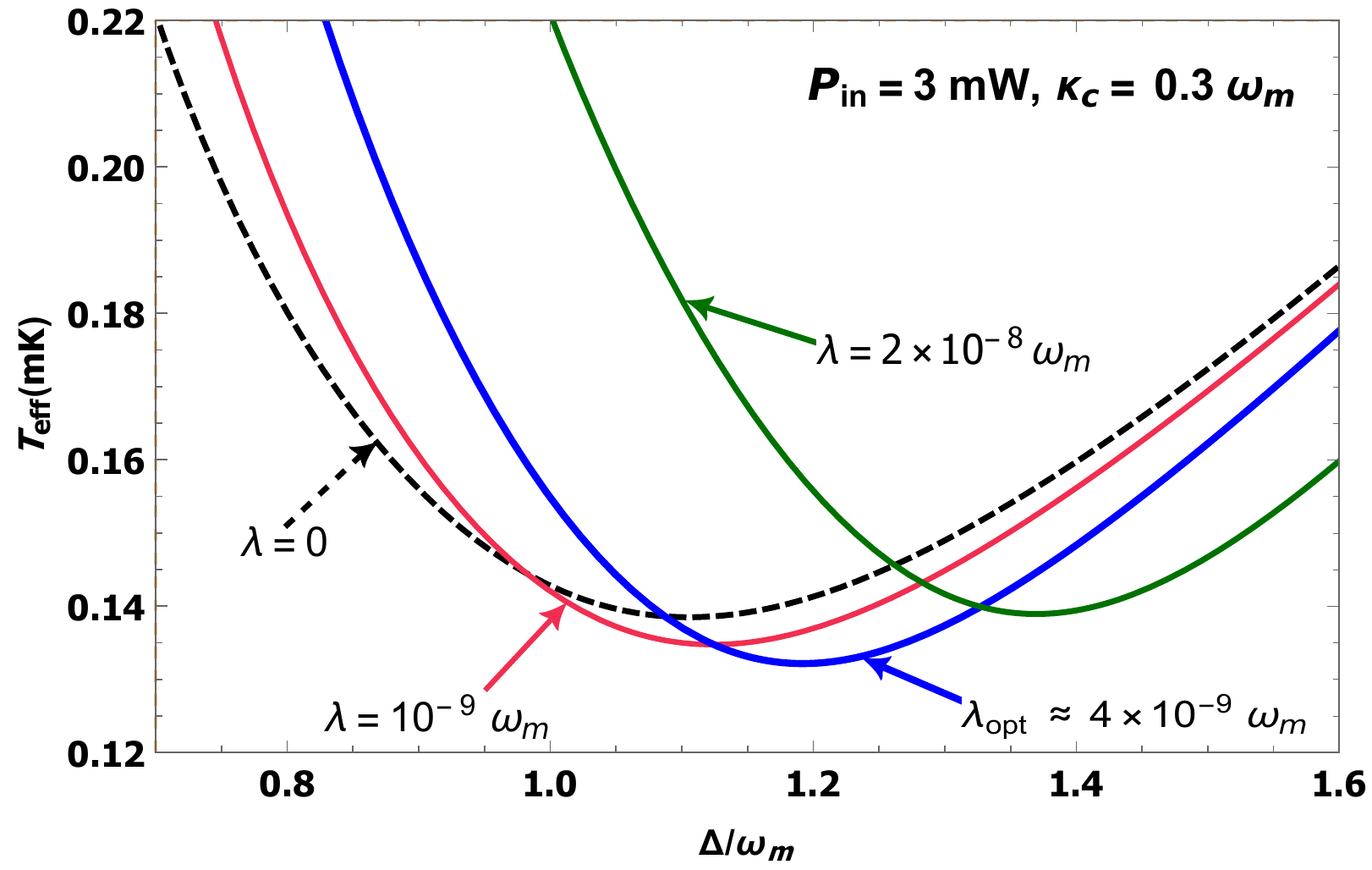}
\caption{ \label{fig6}The effective temperature of the mechanical mode $T_{eff}$ versus the normalized bare detuning $\Delta/\omega_m$ in a bare cavity ($G_0=0$) for $\kappa_c=0.3\omega_m$, $P_{in}=3\text{mW}$,  and different values of the mechanical Duffing anharmonicity strength $\lambda$.  Other parameters are the same as those in \cref{fig5}.}%
\end{figure}
In Fig.~\ref{fig6} we have plotted the effective mode temperature $T_{eff}$ versus the normalized bare detuning $\Delta/\omega_m$ for different values of the Duffing anharmonicity strength $\lambda$. From this figure, it can be found that for some nonzero values of the Duffing parameter the mechanical anharmonicity can result in the reduction of the  minimum attainable value of the effective temperature of the moving mirror. In fact,  there is a certain range of  $\lambda$, which we refer to it as a \textit{mechanical cooling window}, over which the mechanical anharmonicity leads to the reduction of the the effective mode temperature. Also, there is an optimum value of $\lambda$, denoted by $\lambda_{\text{opt}}$ corresponding to $r_{\text{opt}}$, for which the mechanical anharmonicity has the greatest impact on the reduction of the steady-state mean phonon number. To justify this claim, in Fig.~\ref{fig7} we have plotted the effective temperature of the mechanical mode $T_{eff}$ versus the normalized Duffing anharmonicity strength $\lambda/\omega_m$ for different values of the input laser power. In addition,  from this figure it is evident that with increasing $P_{in}$ (which leads to $\eta<1$) the optimum value $\lambda_{opt}$ shifts toward smaller values and the difference between the temperature associated with $\lambda_{\text{opt}}$ and those corresponding to the values of $\lambda$ outside the cooling window including $\lambda=0$ becomes considerable. This temperature difference is due to the presence of the factor $e^{-4r}/\eta$, which we mentioned before.
 \begin{figure}[h!]
 \centering
   \includegraphics[scale=0.48]{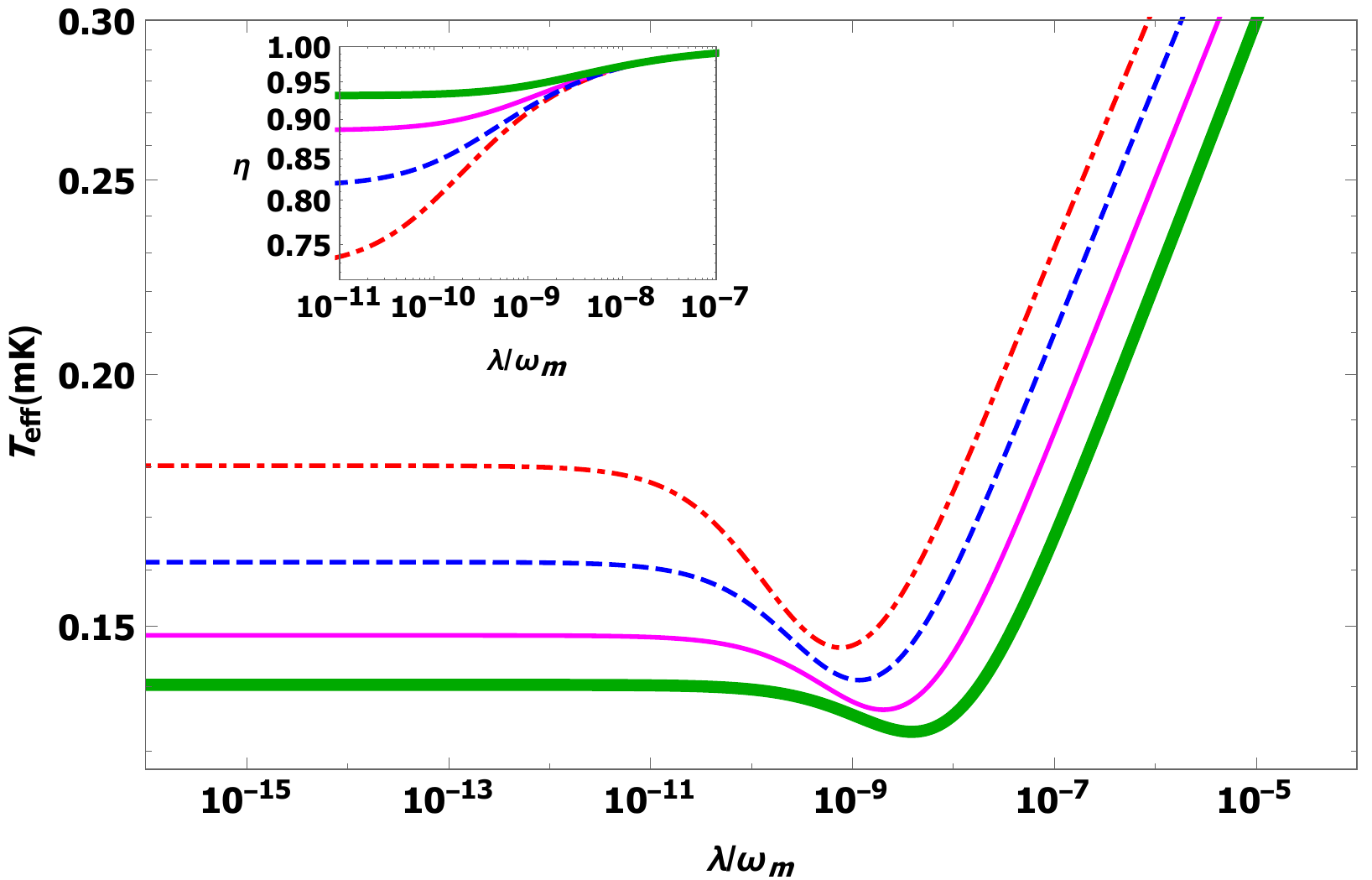}  
\caption{ \label{fig7}The effective temperature of the mechanical mode $T_{eff}$ versus the normalized Duffing anharmonicity strength $\lambda/\omega_m$ for a bare cavity ($G_0=0$) with different values of the input laser power: $P_{in}=3\,\text{mW}$ (thick green line), $P_{in}=5\,\text{mW}$ (thin magenta line), $P_{in}=8\,\text{mW}$ (dashed blue line), $P_{in}=12\,\text{mW}$ (dot dashed red line). The inset figure represents the variation of the stability parameter $\eta$ with the normalized Duffing anharmonicity strength $\lambda/\omega_m$ and the same different values of $P_{in}$. In this figure we have set $\kappa_c=0.3\omega_m$ and $\Delta'=\Omega_m$ in which $\Omega_m$ has been defined just after Eq.~(\ref{rlqle}). Other parameters are the same as those in \cref{fig6}.}%
 \end{figure}
 
\begin{figure}[h!]
 \centering
   \includegraphics[scale=0.5]{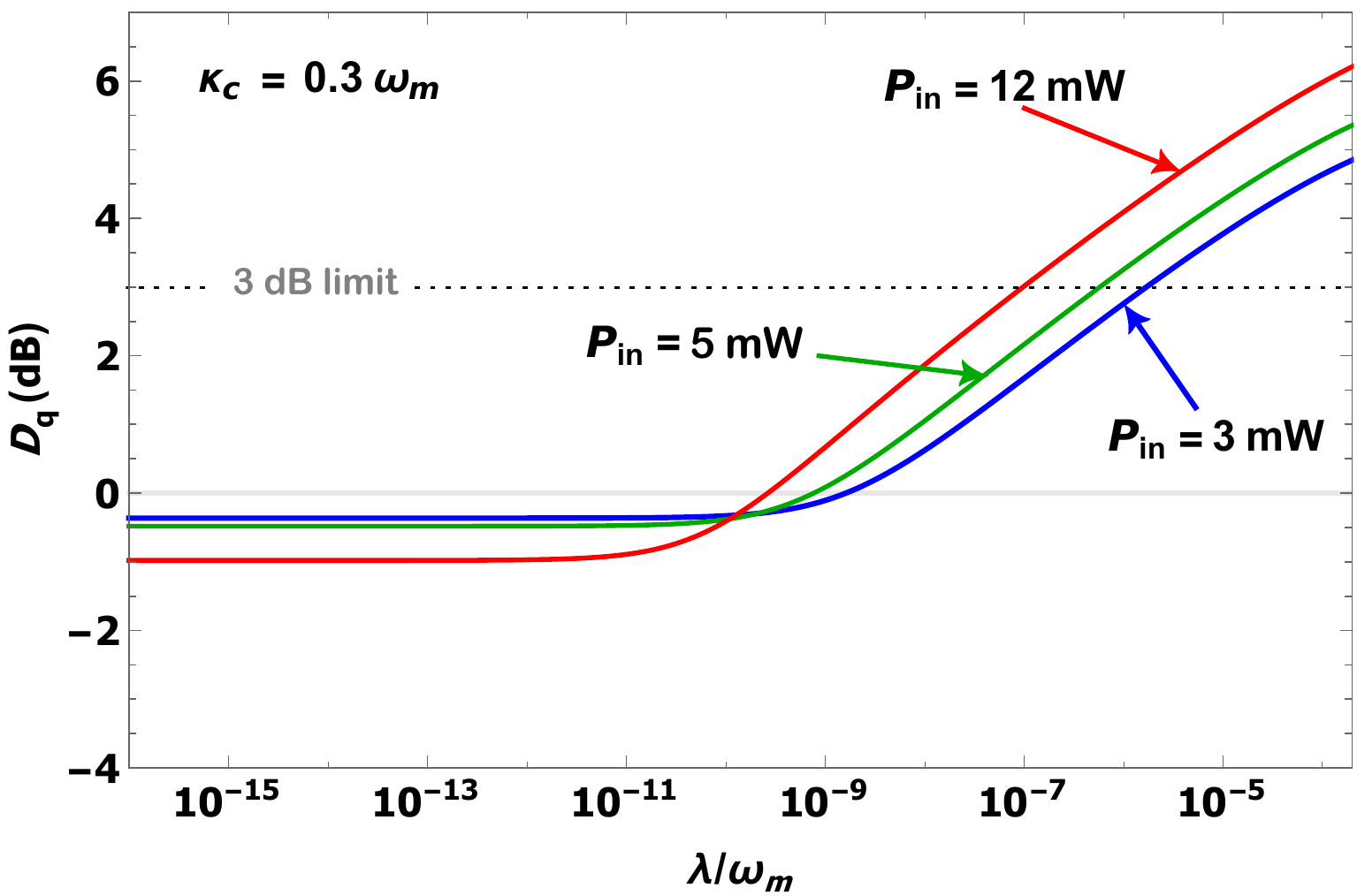}
\caption{ \label{fig8}The steady-state degree of the mechanical squeezing $D_q$ in the dB unit versus the normalized Duffing anharmonicity strength $\lambda/\omega_m$ in a bare cavity ($G_0=0$) for $\kappa_c=0.3\omega_m$  and different values of the  input laser power $P_{in}$.  The dotted line corresponds to the steady-state mechanical squeezing at the standard quantum limit of 3 dB.  In this figure we have set $\Delta'=\Omega_m$ and other parameters are the same as those in \cref{fig7}.}%
\end{figure}

As we found from Fig.~\ref{fig5}(a), with increasing the strength of the mechanical nonlinearity the squeezing of the displacement quadrature fluctuation of the movable mirror increases. Another parameter that can affect the degree of the mechanical squeezing is the input laser power. In Fig. \ref{fig8}, we have plotted the steady-state degree of the mechanical squeezing $D_q$ as a function of the normalized Duffing anharmonicity strength $\lambda/\omega_m$ for different values of the input laser power $P_{in}$ when $G_0=0$. The figure clearly shows that with increasing the input laser power not only the mechanical squeezing starts to be appeared at smaller values of $\lambda$, but also the standard 50\% squeezing ($\equiv$ 3 dB) limit~\cite{o66} can be beaten more strongly allowing more perfect mechanical squeezing. Numerical results show that at the onset of the mechanical squeezing the enhanced Duffing parameter $\Lambda $ approximately equals to the enhanced optomechanical coupling strength $\bar{G}$. In addition, comparison of Figs.~\ref{fig7} and \ref{fig8} reveals that the larger the generated squeezing is, the heater the mechanical motion is.\\ It should be noted that the maximum value of the degree of the mechanical squeezing occurs when $\Delta'\approx\Omega_m$ is established. Therefore, Fig.~\ref{fig8} actually shows the variation of the maximum value of $D_q$ with the normalized Duffing anharmonicity strength $\lambda/\omega_m$.

\begin{figure}[h!!]
 \centering
   \includegraphics[scale=0.47]{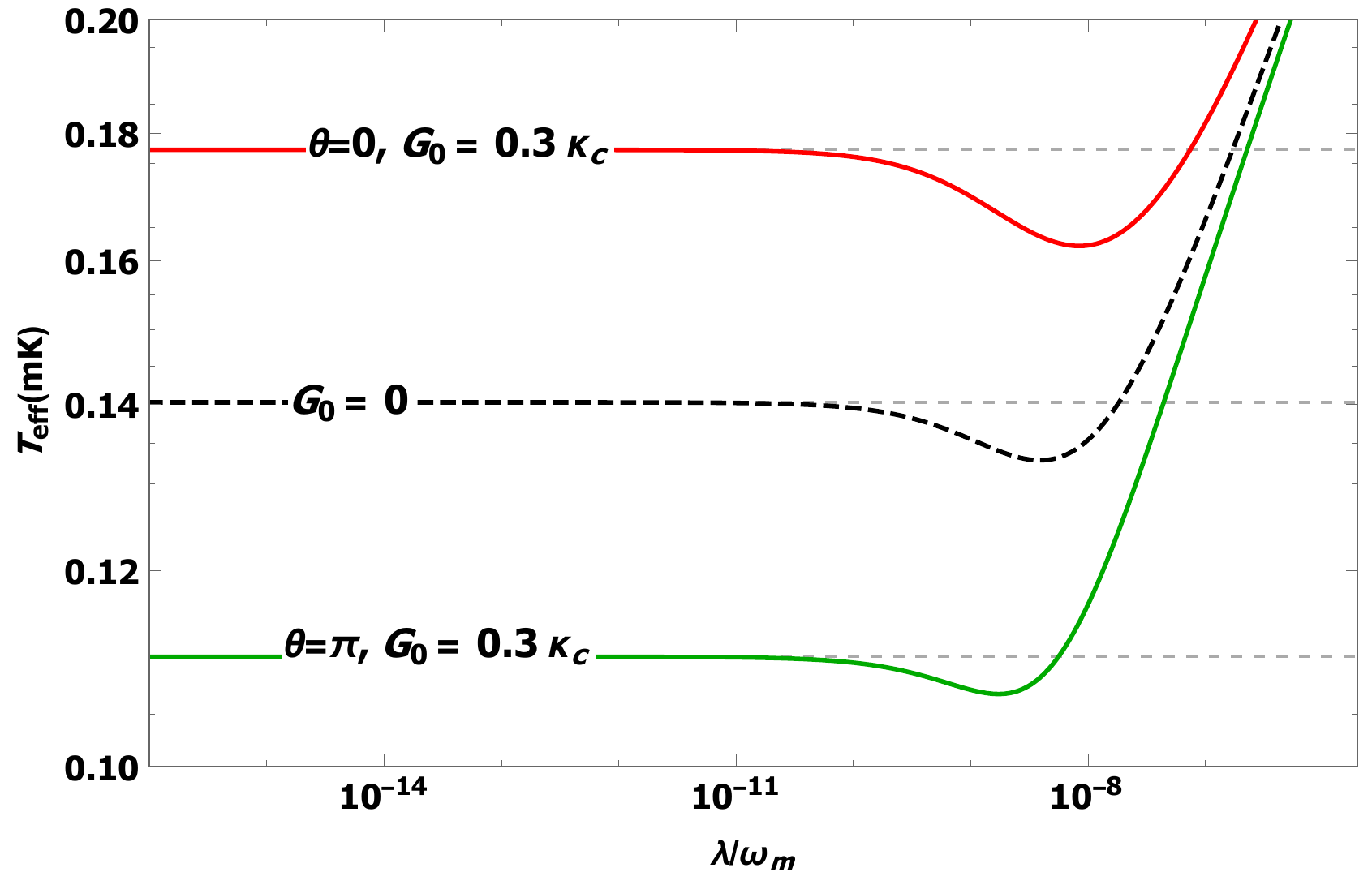} 
\caption{ \label{fig9}The effective temperature $T_{\text{eff}}$  versus the normalized Duffing anharmonicity strength $\lambda/\omega_m$ for different values of the phase $\theta$ when $P_{in}=3\,\text{mW}$, $\kappa_c=0.3\omega_m$, and $G_0=0.3\kappa_c$. In this figure we have set $\Delta'=\Omega_m$ and other parameters are the same as those in Fig.~\ref{fig8}. For comparison purpose, we have shown $T_{\text{eff}}$ for $G_0=0$.}%
\end{figure}
\begin{figure}[h!!]
 \centering
   \includegraphics[scale=0.47]{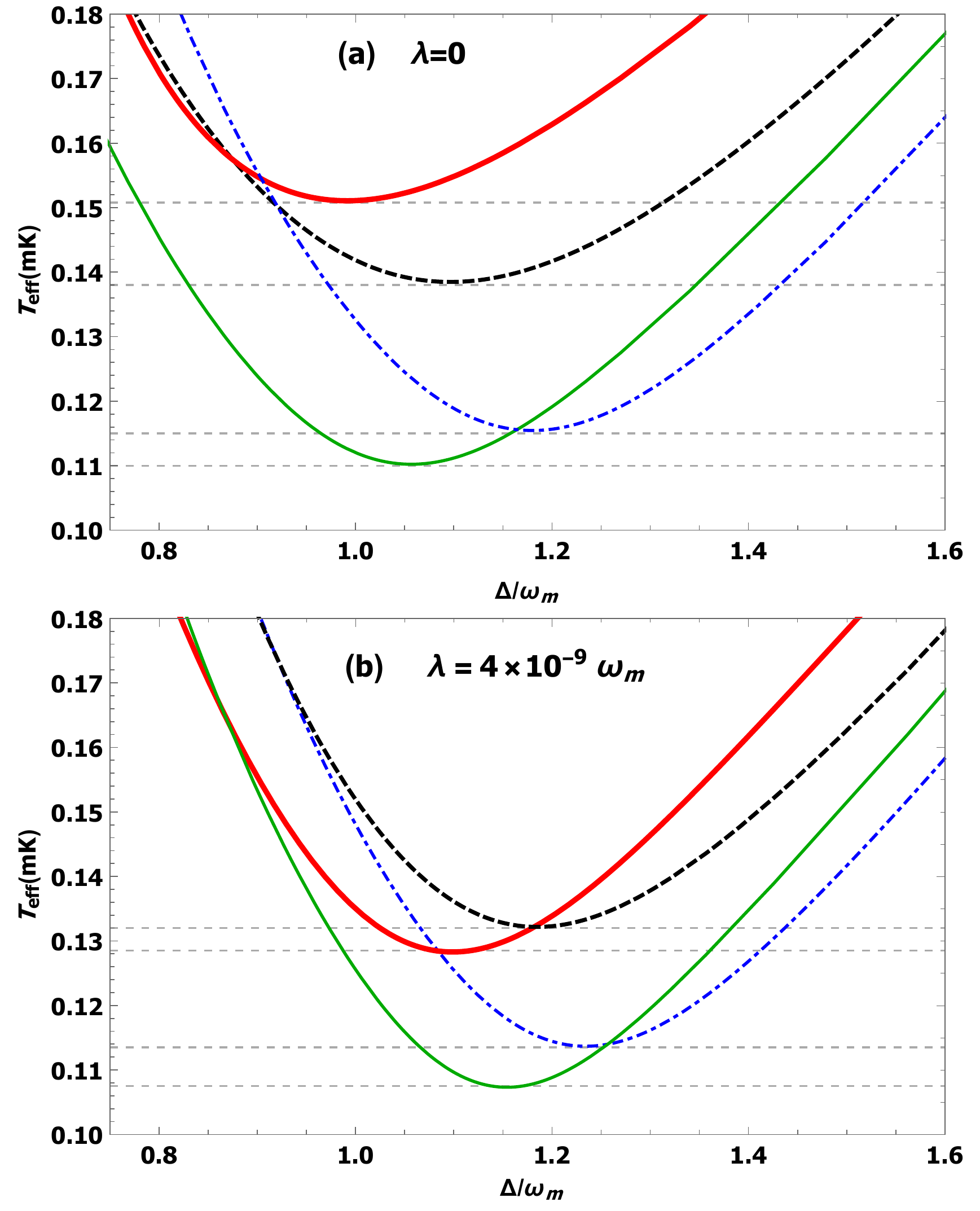} 
\caption{ \label{fig10}The effective temperature $T_{\text{eff}}$ versus the normalized bare detuning $\Delta/\omega_m$ for two different values of the anharmonicity parameter: (a) $\lambda=0$, (b) $\lambda=4\times10^{-9}\omega_m$; with $P_{in}=3\,\text{mW}$, and different values of ($G_0$, $\theta$): ($0.3\kappa_c$,$\pi$) (thin green lines), ($0.3\kappa_c$,$1.3\pi$)(dot dashed blue lines), ($0.6\kappa_c$,$0.71\pi$)(thick red lines), and $G_0=0$(dashed black lines). In this figure we have set  $\Delta'=\Omega_m$. Other parameters are the same as those in Fig.~\ref{fig9}.}%
\end{figure}
Until now, we have investigated the effects of the Duffing nonlinearity on the ground-state cooling and quadrature  squeezing of the mechanical mode in the absence of the OPA. Now, we focus our attention on the role of the OPA in the cooling and quadrature squeezing of the Duffing-like mechanical oscillator. In Fig.~\ref{fig9} we have plotted the effective temperature of the vibrational mode of the moving mirror, $T_{\text{eff}}$, versus the normalized Duffing anharmonicity strength $\lambda/\omega_m$ for  $G_0=0.3\kappa_c$ and various values of the parametric phase $\theta$ in the good-cavity limit. From this figure it is clear that depending on the value of $\theta$, the OPA can manifest itself by narrowing or broadening the width of the cooling window, that is the range of $\lambda$ over which the mechanical anharmonicity can be exploited to cool-down the mechanical motion. Moreover, by proper choice of the phase $\theta$, it may be possible to reduce considerably the minimum attainable value of $T_{\text{eff}}$. Figures \ref{fig10}(a) and \ref{fig10}(b) show the variation of the effective temperature $T_{\text{eff}}$ with $\Delta/\omega_m$ and properties of the nonlinear gain medium ($G_0$, $\theta$) for $\lambda=0$ and $\lambda=4\times10^{-9}\omega_m$, respectively. From these figures (except for the thick red lines) it is clear that, as expected, the presence of the OPA leads to a decrease in the effective mode temperature $T_{\text{eff}}$ both in the absence and in the presence of the mechanical nonlinearity. But for non-zero $\lambda$, it leads to less effective temperature. Comparing the thick red lines in (a) and (b) which are correspond to $G_0=0.6\kappa_c$ and $\theta=0.71\pi$, it is understood that for some values of ($G_0$, $\theta$), which result in an increase in the effective temperature of the harmonic oscillator, the Duffing mechanical anharmonicity can neutralize and even reverse the effect of the OPA. In addition, the red line in panel (a) can be considered as an example that violates Eq.~(\ref{nef2}) due to the contribution of the non-zero value of the thermal phonon number, although  the low temperature environment exists ($\kappa/(\gamma \bar{n}_m)\approx 5821$).

\begin{figure}[h!]
 \centering
   \includegraphics[scale=0.461]{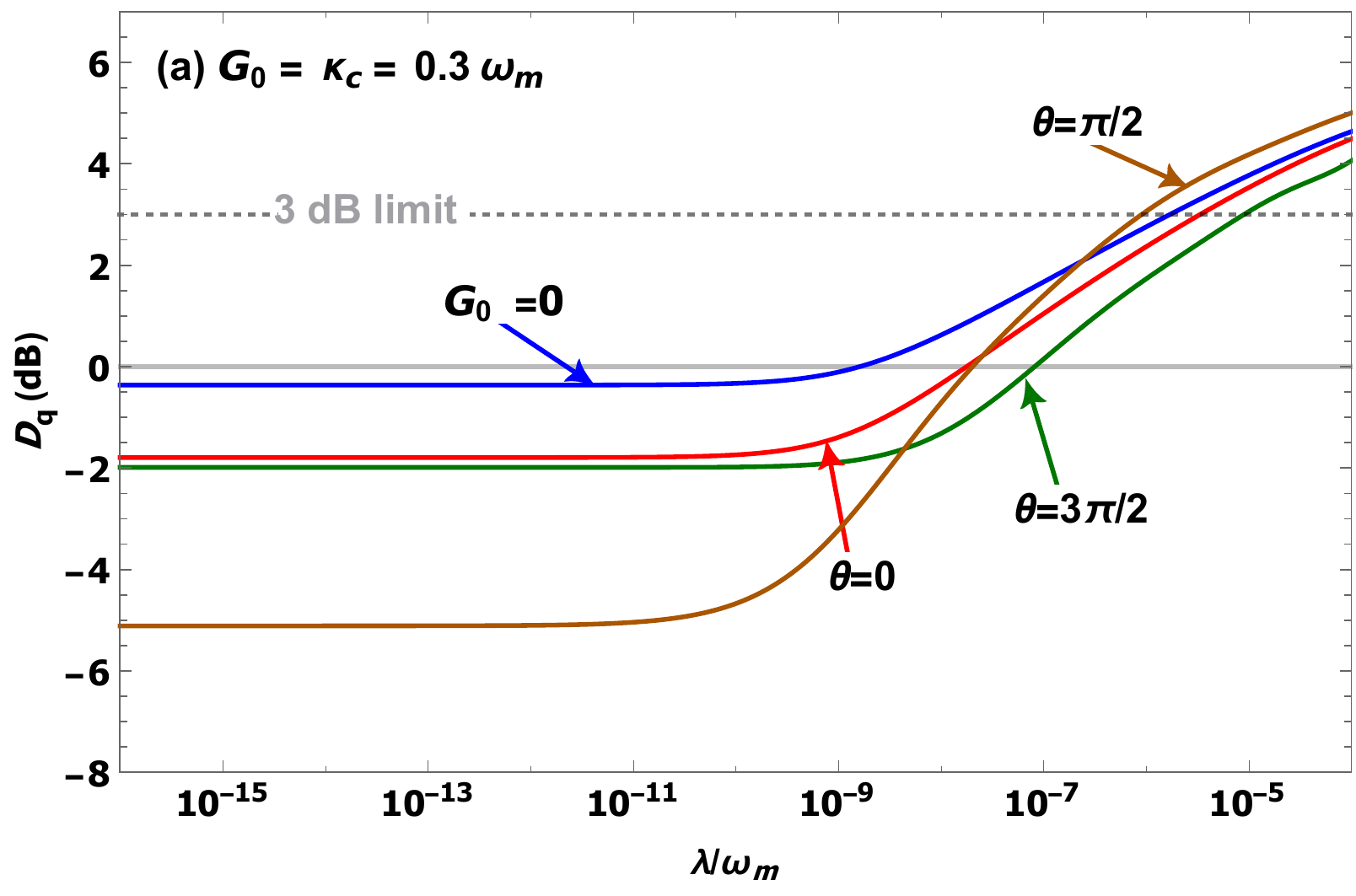} \par \includegraphics[scale=0.465]{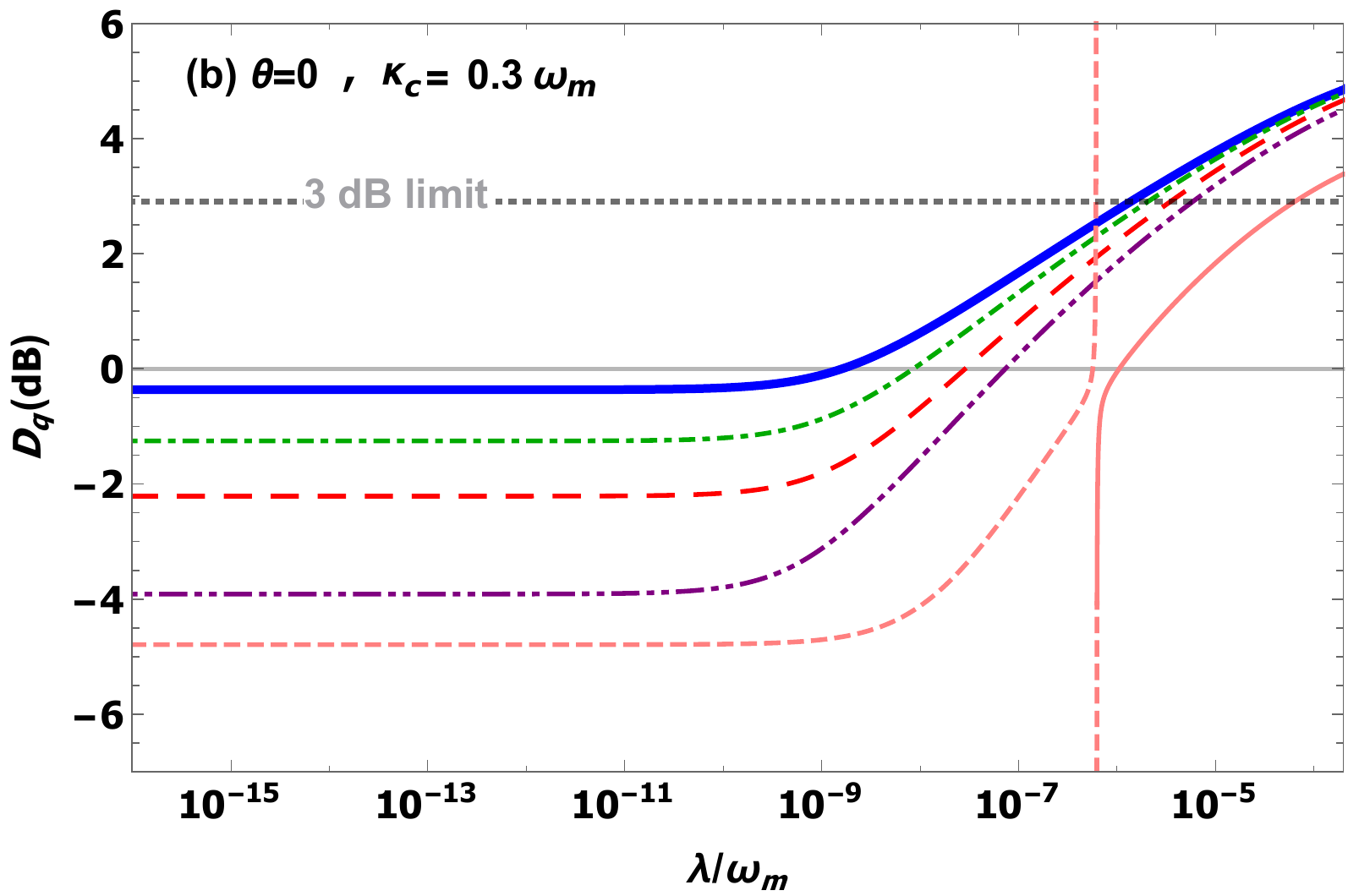}
\caption{ \label{fig11}The steady-state degree of the mechanical squeezing $D_q$ in the dB unit versus the normalized Duffing anharmonicity strength $\lambda/\omega_m$ with $\kappa_c=0.3\omega_m$ for (a) $G_0=\kappa_c$ and different values of  $\theta$, and (b) $\theta=0$ and different values of $G_0$: $G_0=0$ (thick blue line), $G_0=0.7\kappa_c$ (dot dashed green line), $G_0=1.2\kappa_c$ (long- dashed red line), $G_0=1.6\kappa_c$ (double-dot dashed purple line), $G_0=2.8\kappa_c$ (thin pink line for stable region and dashed pink line for unstable region). The  3 dB limit of squeezing has also been shown. Here we have set $\Delta'=\Omega_m$ and other parameters are the same as those in Fig.~\ref{fig10}. In panel (a) $D_q$ for $G_0=0$ has been plotted for comparison.}%
    \end{figure}

Finally, we examine the dependence of the steady-state degree of the mechanical squeezing $D_q$ on the nonlinear gain $G_0$ and the parametric phase $\theta$.  In Fig.~\ref{fig11}(a) we have plotted the degree of the mechanical squeezing $D_q$ at steady state versus the normalized Duffing anharmonicity strength $\lambda/\omega_m$ for $G_0=\kappa_c=0.3\omega_m$ and different values of $\theta$. This figure clearly shows that the OPA causes the onset of the mechanical squeezing  to be shifted toward larger values of $\lambda$ and,  depending on the values of $\theta$ and $\lambda$, the OPA can enhance or weaken the steady-state mechanical squeezing. As mentioned above, the starting point of the mechanical squeezing is when $\Lambda\approx \bar{G}$. The OPA, by changing both $\Lambda$ and $\bar{G}$, can alter the degree of the mechanical squeezing as well as its onset.
    \begin{figure}[h!!]
 \centering
   \includegraphics[scale=0.445]{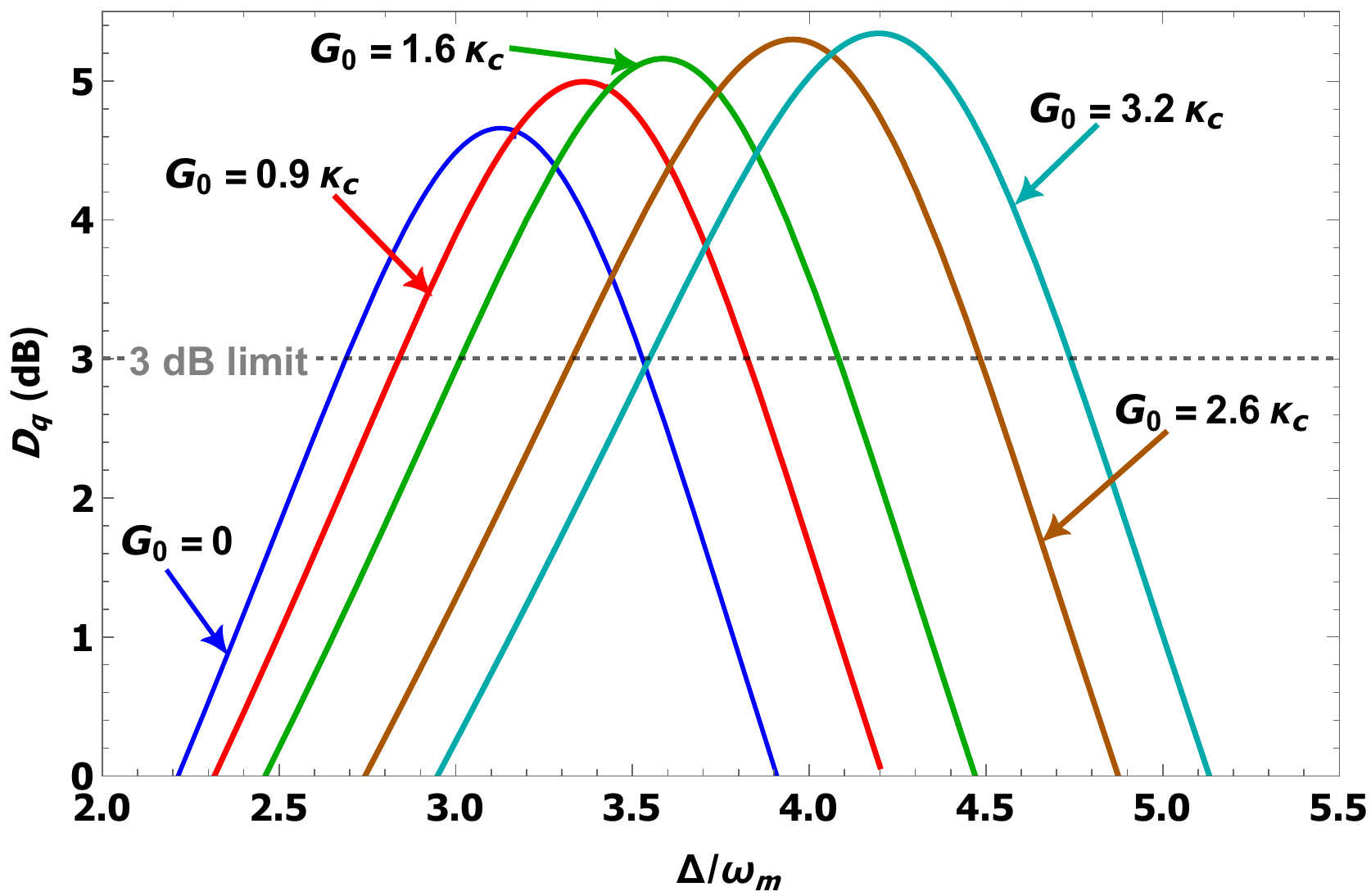} 
\caption{ \label{fig12}The steady-state degree of the mechanical squeezing $D_q$ in the dB unit versus the normalized bare detuning $\Delta/\omega_m$ for different values of $G_0$ with $\lambda=10^{-4}\omega_m$, $P_{in}=3\,\text{mW}$, $\kappa_c=0.3\omega_m$,  and $\theta=\pi/2$. The figure has been plotted in a range of $\Delta/\omega_m$ for which $\beta_s\geq40$. The maximum value of the degree of the mechanical squeezing occurs when $\Delta'=\Omega_m$. Other parameters are the same as those in Fig.~\ref{fig11}.}%
    \end{figure}
Figure \ref{fig11}(b) shows the steady-state degree of the mechanical squeezing $D_q$ as a function of the normalized Duffing anharmonicity strength $\lambda/\omega_m$ for various values of the nonlinear gain $G_0$ with $\theta=0$. As can be seen, with increasing the value of $G_0$, the onset of the mechanical squeezing happens at larger values of $\lambda$ and the degree of squeezing decreases. In Fig.~\ref{fig12}, we have plotted the steady-state degree of the mechanical squeezing $D_q$ versus the bare cavity detuning $\Delta/\omega_m$ for $\lambda=10^{-4}\omega_m$, $\theta=\pi/2$, and different values of $G_0$. As is seen, with increasing  $G_0$  the maximum of the degree of the mechanical squeezing is shifted to higher values of  the bare detuning while being amplified, and the range of $\Delta/\omega_m$ over which the displacement quadrature of the moving mirror is squeezed beyond the 3 dB limit being wider.

\section{\label{sec:4th}Conclusions}
In conclusion, we have studied theoretically an optomechanical cavity with a Duffing-like movable mirror which contains a degenerate OPA. We have investigated the multistability in the steady-state mean intracavity photon number and  mechanical oscillation amplitude by providing analytical expressions for the critical values of the system parameters corresponding to the emergence of the multistable behavior. We have also explored the roles of the Duffing anharmonicity and the gain nonlinearity in the ground-state cooling of the movable mirror and the steady-state mechanical squeezing. 

We have found  that the stiffening Duffing mechanical anharmonicity reduces the width of the multistability region by suppressing the mechanical oscillation amplitude as well as increasing the critical values of the system parameters, while the parametric optical nonlinearity can be exploited to drive the system toward multistability region by increasing the mechanical oscillation amplitude and decreasing the critical values of the system parameters. Moreover, we have found that in the absence of the OPA there is a certain range of the Duffing anharmonicity strength, the so-called mechanical cooling window, over which the mechanical anharmonicity leads to the cooling of the mechanical motion. In addition, as the Duffing mechanical anharmonicity strength reaches a threshold value the mechanical squeezing beyond the 3 dB limit can be achieved. The results reveal that  the effective temperature of the mechanical motion,  the onset of the mechanical squeezing, and the threshold value of the Duffing mechanical anharmonicity strength to achieve squeezing exceeding the 3 dB limit depend on the system parameters such as the input laser power, the parametric nonlinearity, and the phase of the field driving the OPA. In particular, we have shown that the Duffing mechanical anharmonicity has significant effect on the ground-state cooling of the mechanical motion. Furthermore, by choosing properly the parametric phase as well as the nonlinear gain the mechanical anharmonicity-induced cooling of the mechanical motion can be greatly enhanced and  the strong steady-state mechanical squeezing beyond the 3 dB limit can be reached.

\section*{Appendix: Critical values of the system parameters corresponding to the emergence of multistability}

\setcounter{equation}{0}
\renewcommand{\theequation}{A\arabic{equation}}
In this appendix, we derive the critical values of the system parameters corresponding to the emergence of the multistable behavior. To determine the critical power, we follow here the approach outlined in Ref.~\cite{o67}. In fact the critical points correspond to vertical tangencies (infinite slope) of the response curve (amplitude versus detuning curve). 

Equations (\ref{bs}) and (\ref{aas}) lead to following fifth-order equation for the mechanical amplitude:
\begin{small}\begin{align}\label{quintic}
&64 g^2 \lambda \beta_s^5\,-\, 64g \lambda\,(\Delta-2G_0 \sin\theta)\beta_s ^4\,+\,4\Big \{ g^2(\omega_m+12\lambda) \nonumber\\&+4\lambda[(\kappa_c-2G_0 \cos\theta)^2+ (\Delta-2G_0 \sin\theta)^2] \Big\}\beta_s^3 \nonumber\\ &-\, 4g(\,\omega_m+12 \lambda )(\Delta-2G_0 \sin\theta)\beta_s^2\,\nonumber\\&+\,(\omega_m+12\lambda)[(\Delta-2G_0 \sin\theta)^2+(\kappa_c-2G_0 \cos\theta)^2]\beta_s\,\nonumber\\&\hspace*{5cm}-\, g \varepsilon^2\,=\,0.
\end{align}\end{small}
The first point that can be deduced from Eq.~(\ref{quintic}) is that there is a possibility of multistability in the stationary response of the mechanical resonator. Using Descartes's rule of signs~\cite{o68}, one can count the number of real positive zeros of Eq.~(\ref{quintic}).
From Eq.~(\ref{quintic}), with the assumption $\lambda>0$, it is clear that  the coefficients of $\beta_s^5$, $\beta_s^3$ and $\beta_s$ are always positive and $-\, g \varepsilon^2$ is always negative. The coefficients of $\beta_s^4$ and $\beta_s^2$ have the same sign, so if they have positive values then Eq.~(\ref{quintic}) has only one real root and  if they have negative values Eq.~(\ref{quintic}) has 5 or 3 or 1 real roots. Here, we are looking for conditions under which Eq.~(\ref{quintic}) has more than one real root.

Since $\beta_s$ is a function of $d_0=\Delta-2G_0 \sin\theta$,  differentiating Eq.~(\ref{quintic}) with respect to $d_0$ results in
\begin{small}\begin{eqnarray}\label{dff}
\partial_{d_0}\beta_s\,=\,\frac{P[d_0]}{Q[d_0]},
\end{eqnarray}\end{small}
where
\begin{small}\begin{align}
P[d_0]=&-2\beta_s\,(d_0-2g \beta_s)\,(\omega_m+12\lambda+16\lambda\beta_s^2),\\
Q[d_0]=&(\omega_m+12\lambda)\{ d_0^2+(\kappa_c-2G_0 \cos\theta)^2\nonumber\\&-8g\,d_0\, \beta_s\,+12g^2 \beta_s^2\}+\,48\lambda[d_0^2+(\kappa_c-2G_0 \cos\theta)^2]\beta_s^2\nonumber\\&\hspace*{1cm}-256g \lambda \,d_0 \beta_s^3 +320 g^2 \lambda \beta_s^4.
\end{align}\end{small}
By equating the denominator of Eq.~(\ref{dff}) to zero and solving the resultant quadratic equation for $\Delta$ one can obtain 
\begin{small}\begin{align}\label{qdis}
\hspace*{-0.3cm}\Delta\,=\,2G_0 \sin\theta+\frac{128 \beta_s^3 g \lambda + 4 \beta_s g (\omega_m+12 \lambda )\pm \sqrt{\Delta_{quad}}}{12 (1 + 4 \beta_s^2) \lambda + \omega_m},
\end{align}\end{small}
in which $\Delta_{quad} $ defined by
\begin{small}\begin{eqnarray}\label{qqdis}
&&\Delta_{quad}=R_3 \beta_s^6 \,+\,R_2 \beta_s^4\,+\,R_1 \beta_s^2 +R_0,
\end{eqnarray}\end{small}
is the discriminant of the equation $Q[d_0]=0$ with $\Delta$ as the variable. In Eq.~(\ref{qqdis}) $R_j$'s are defined as
\begin{small}\begin{eqnarray}
&&\hspace{-1.5cm}R_3=1024 g^2 \lambda^2,
\\&&\hspace{-1.5cm}R_2=128 \lambda [g^2 (\omega_m+12 \lambda)-18\lambda (\kappa_c-2G_0 \cos\theta)^2  ],
\\&&\hspace{-1.5cm}R_1=4(\omega_m+12 \lambda) [g^2(\omega_m+12 \lambda)\nonumber
\\&&\hspace*{1cm}-24\lambda (\kappa_c-2G_0 \cos\theta)^2 ],\\&&\hspace{-1.5cm}R_0=- (\kappa_c-2G_0 \cos\theta)^2 ( \omega_m+12 \lambda )^2.
\end{eqnarray}\end{small}
Just at the critical point, $\Delta_{quad}$ must be zero and in multistability region we have $\Delta_{quad}>0$. Therefore we first determine the roots of the discriminant, namely we solve equation $\Delta_{quad}=0$ which is a cubic equation for $\beta_s^2$ and its discriminant, denoted by $\Delta_{cub}$, provides information about roots:\begin{small}
\begin{equation}
\Delta_{cub}=(\frac{R_1}{3R_3}-\frac{R_2^2}{9R_3^2})^3+(\frac{R_0}{2R_3}-\frac{R_1R_2}{6R_3^2}+\frac{R_2^3}{27R_3^3})^2.
\end{equation}\end{small}
In this way, we find that 
\begin{enumerate}
\item $\Delta_{quad}=0$ has one real root and two conjugate imaginary roots and $\Delta$ has one critical value if $ \Delta_{cub}>0$;
\item  $\Delta_{quad}=0$ has three real roots of which at least two are equal and $\Delta$ has one or two critical values if $ \Delta_{cub}=0$;
\item $\Delta_{quad}=0$ has three unequal real roots and $\Delta$ has three different critical values if $ \Delta_{cub}<0$.
\end{enumerate}
By assuming $\lambda>0$, a straightforward calculation shows that for our problem we always have $ \Delta_{cub}>0$ and the only real root of the equation $\Delta_{quad}=0$ is \begin{small}
\begin{align}
&\beta_s^{crit}=\Big[-\frac{R_2}{3R_3}\,+\,\sqrt[3]{-(\frac{R_0}{2R_3}-\frac{R_1R_2}{6R_3^2}+\frac{R_2^3}{27R_3^3})+\sqrt{\Delta_{cub}}}\nonumber\\& \,-\,(\frac{R_1}{3R_3}-\frac{R_2^2}{9R_3^2})\frac{1}{\sqrt[3]{-(\frac{R_0}{2R_3}-\frac{R_1R_2}{6R_3^2}+\frac{R_2^3}{27R_3^3})+\sqrt{\Delta_{cub}}}}\Big]^{\frac{1}{2}}.
\end{align}\end{small}
Finally, by replacing $\beta_s$ and $\Delta$ in Eq. (\ref{quintic}), respectively, with $\beta_s^{crit}$ and $\Delta^{crit}$,\begin{small}
\begin{equation}
\Delta^{crit}\,=\,2G_0 \sin\theta+\frac{128\beta_s^{crit^3} g \lambda + 4 \beta_s^{crit} g (\omega_m+12 \lambda )}{12 (1 + 4 \beta_s^{crit^2}) \lambda + \omega_m},
\end{equation}\end{small}and solving the resulting equation for $P_{in}$ we find the critical value for the input laser power $P_{in}^{crit}$. It should be noted that in order to enter the multistability region  the conditions $\beta_s\,>\,\beta_s^{crit}$, $\Delta\,>\,\Delta^{crit}$ and $P_{in}\,>\,P_{in}^{crit}$ must be fulfilled simultaneously.

In the case of harmonic oscillator, we set parameter $\lambda$ in Eq.~(\ref{quintic})  equal to zero and use the same method as explained above to obtain the following critical values:\begin{small}
\begin{align}
& (\beta_{s}^{crit})_{\lambda=0}=\Big|\frac{\kappa_c\,-\,2G_0\,\cos\theta}{2\,g}\Big|,\\& (\Delta^{crit})_{\lambda=0}\,=\,2G_0\,\sin\theta\,+\,4g (\beta_{s}^{crit})_{\lambda=0},\\&(P_{in}^{crit})_{\lambda=0}=\frac{\hbar \omega_L \omega_m}{g\,\kappa_c}(\kappa_c\,-\,2G_0\,\cos\theta)^2\,(\beta_{s}^{crit})_{\lambda=0}.
\end{align}\end{small}

\end{document}